\pdfoutput=1
\documentclass[a4paper,UKenglish,cleveref, autoref, thm-restate]{lipics-v2021}

\newcommand{\overbar}[1]{\mkern 1.5mu\overline{\mkern-1.5mu#1\mkern-1.5mu}\mkern 1.5mu}

\newcommand{\one}{1}
\newcommand{\two}{P2}
\newcommand{\three}{P3}

\newcommand{\fixed}{Pm}

\newcommand{\size}{size}
\newcommand{\any}{any}
\newcommand{\rej}{Rej\leq Q}
\newcommand{\rj}{r_j}
\renewcommand{\dj}{C_j\leq d_j}

\newcommand{\cmax}{C_{\max}}
\newcommand{\lmax}{L_{\max}}
\newcommand{\tmax}{T_{\max}}
\newcommand{\fmax}{F_{\max}}
\newcommand{\sumc}{\sum C_j}
\newcommand{\sumu}{\sum U_j}

\newcommand{\sumt}{\sum T_j}

\newcommand{\sumwu}{\sum w_jU_j}

\newcommand{\sumwt}{\sum w_jT_j}
\newcommand{\sumwc}{\sum w_jC_j}
\newcommand{\sumwf}{\sum w_jF_j}

\newcommand{\tf}[3]{${#1|#2|#3}$}

\newcommand{\SSS}{\textsc{Subset Sum}\xspace}
\newcommand{\KS}{\textsc{Knapsack}\xspace}
\newcommand{\PART}{\textsc{Partition}\xspace}

\newcommand{\Oh}[1]{\mathcal{O}\left(#1\right)}
\newcommand{\oh}[1]{o\left(#1\right)}
\newcommand{\Ohtilde}[1]{\tilde{\mathcal{O}}\left(#1\right)}

\newcommand{\etal}{et~al.\xspace}

  \newenvironment{myproblem}%
{%
  \leavevmode\nobreak\par
	\begin{list}%
		{}%
		{%
			\def\labelstyle{\itshape}
			\setlength{\topsep}{0pt}%
			\settowidth{\labelwidth}{\labelstyle Parameter:}%
			\setlength{\leftmargin}{\labelwidth}%
			\addtolength{\leftmargin}{\labelsep}%
			\setlength{\itemsep}{0pt}%
			\setlength{\parsep}{0pt}%
		}%
		\def\instance{\item[\labelstyle Instance:]}%
    \def\task{\item[\labelstyle Task:]}%
	}%
	{%
	\end{list}%
}
\usepackage{acro}
\usepackage{xspace}
\usepackage{bm}
\usepackage{csquotes}

\makeatletter
\def\hlinewd#1{%
\noalign{\ifnum0=`}\fi\hrule \@height #1 \futurelet
\reserved@a\@xhline}
\makeatother

\newtheorem{problem}[theorem]{Problem}

\newcommand{\OPT}{\operatorname{\text{\textsc{opt}}}}

\usepackage{tikz}
\usetikzlibrary{positioning}
\usetikzlibrary{calc}

\usepackage{mathtools}
\usepackage{amssymb}
\usepackage{graphicx}
\usepackage{makecell}
\usepackage{xltabular}

\newcommand{\eps}{\varepsilon}

\DeclarePairedDelimiter\set{\lbrace}{\rbrace}
\DeclarePairedDelimiterX\sett[2]{\lbrace}{\rbrace}{ #1 \,\delimsize| \,\mathopen{} #2 }

\title{On the Complexity of Scheduling Problems With a Fixed Number of Parallel Identical Machines}

\titlerunning{On the Complexity of Scheduling Problems}

\author{Klaus Jansen}{Kiel University, Department of Computer Science,  Germany}{kj@informatik.uni-kiel.de}{}{}
\author{Kai Kahler}{Kiel University, Department of Computer Science,  Germany}{kka@informatik.uni-kiel.de}{}{}

\funding{Both authors are supported by the German Research Foundation (DFG) project JA 612/25-1.}



\authorrunning{K. Jansen and K. Kahler}

\Copyright{Klaus Jansen, and Kai Kahler}

\supplement{}

\acknowledgements{The authors wish to thank Sebastian Berndt, Max Deppert, Sören Domrös, Lena Grimm, Leonie Krull, Marten Maack, Niklas Rentz and anonymous reviewers for very helpful comments and ideas.}

\nolinenumbers 

\hideLIPIcs  

\EventEditors{John Q. Open and Joan R. Access}
\EventNoEds{2}
\EventLongTitle{42nd Conference on Very Important Topics (CVIT 2016)}
\EventShortTitle{CVIT 2016}
\EventAcronym{CVIT}
\EventYear{2016}
\EventDate{December 24--27, 2016}
\EventLocation{Little Whinging, United Kingdom}
\EventLogo{}
\SeriesVolume{42}
\ArticleNo{23}

\newif\iflongversion

\longversiontrue
\begin{document}

\maketitle

\begin{abstract}
In parallel machine scheduling, we are given a set of jobs, together with a number of machines and our goal is to decide for each job, when and on which machine(s) it should be scheduled in order to minimize some objective function. Different machine models, job characteristics and objective functions result in a multitude of scheduling problems and many of them are NP-hard, even for a fixed number of identical machines. In this work, we give conditional running time lower bounds for a large number of scheduling problems, indicating the optimality of some classical algorithms. Most notably, we show that the algorithm by Lawler and Moore for \tf{\one}{}{\sumwu} and \tf{\fixed}{}{\cmax}, as well as the algorithm by Lee and Uzsoy for \tf{\two}{}{\sumwc} are probably optimal. There is still small room for improvement for the \tf{\one}{\rej}{\sumwu} algorithm by Zhang \etal, the algorithm for \tf{\one}{}{\sumt} by Lawler and the FPTAS for \tf{\one}{}{\sumwu} by Gens and Levner. We also give a lower bound for \tf{\two}{\any}{\cmax} and improve the dynamic program by Du and Leung from $\Oh{nP^2}$ to $\Oh{nP}$, matching this new lower bound. Here, $P$ is the sum of all processing times. The same idea also improves the algorithm for \tf{\three}{\any}{\cmax} by Du and Leung from $\Oh{nP^5}$ to $\Oh{nP^2}$. While our results suggest the optimality of some classical algorithms, they also motivate future research in cases where the best known algorithms do not quite match the lower bounds.
\end{abstract}
\section{Introduction}
\label{sec:introduction}
Consider the problem of working on multiple research papers. Each paper $j$ has to go to some specific journal or conference and thus has a given due date $d_j$. Some papers might be more important than others, so each one has a weight $w_j$. In order to not get distracted, we may only work on one paper at a time and this work may not be interrupted. If a paper does not meet its due date, it is not important by how much it misses it; it is either late or on time. If it is late, we must pay its weight $w_j$. In the literature, this problem is known as \tf{\one}{}{\sumwu} and it is one of Karp's original 21 NP-hard problems~\cite{Karp72}. The naming of \tf{\one}{}{\sumwu} and the problems referred to in the abstract will become clear when we review the three-field notation by Graham \etal~\cite{GLLK1979} in \cref{sec:preliminaries}. Even when restricted to a fixed number of identical machines, many combinations of job characteristics and objective functions lead to NP-hard problems. For this reason, a lot of effort has been put towards finding either pseudo-polynomial exact or polynomial approximation algorithms. Sticking to our problem \tf{\one}{}{\sumwu}, where we aim to minimize the weighted number of late jobs on a single machine, there are e.g. an $\Oh{nW}$ algorithm by Lawler and Moore~\cite{LM69} and an FPTAS by Gens and Levner~\cite{GL81}. Here, $W$ is the sum of all weights $w_j$ and $n$ is the number of jobs.

In recent years, research regarding scheduling has made its way towards parameterized and fine-grained complexity (see e.g.~\cite{ABHS22,HMS22,KK18,MB18,MW14}), where one goal is to identify parameters that make a problem difficult to solve. If those parameters are assumed to be small, parameterized algorithms can be very efficient. Similarly, one may consider parameters like the total processing time $P$ and examine how fast algorithms can be in terms of these parameters, while maintaining a sub-exponential dependency on $n$. That is our main goal in this work. Most of our lower bounds follow from a lower bound for \SSS:
\begin{problem}{\SSS}
\begin{myproblem}
\instance Items $a_{1},\ldots,a_{n}\in\mathbb{N}$, integer target $T\in\mathbb{N}$.
\task Decide whether there is a subset $S\subseteq [n]$ such that $\sum_{i\in S}a_i = T$.
\end{myproblem}
\end{problem}
Fine-grained running time lower bounds are often based on the Exponential Time Hypothesis (ETH) or the Strong Exponential Time Hypothesis (SETH). Intuitively, the ETH conjectures that \textsc{3-Sat} cannot be solved in sub-exponential time and the SETH conjectures that the trivial running time of $\Oh{2^{n}}$ is optimal for \textsc{$k$-Sat}, if $k$ tends to infinity. For details, see the original publication by Impagliazzo and Paturi~\cite{IP99}.
A few years ago, Abboud et al. gave a beautiful reduction from \textsc{$k$-Sat} to \SSS~\cite{ABHS19}. Previous results based on the ETH excluded $2^{o(n)}T^{o(1)}$-time algorithms~\cite{JLL16}, while this new result based on the SETH suggests that we cannot even achieve $\Oh{2^{\delta n}T^{1-\eps}}$:
\begin{theorem}[Abboud \etal~\cite{ABHS19}]
\label{thm:exact_subsetsum}
For every $\eps>0$, there is a $\delta>0$ such that \SSS cannot be solved in time $\Oh{2^{\delta n} T^{1-\eps}}$, unless the SETH fails.\footnote{Though it might seem unintuitive at first, it is not required that $\eps<1$.}
\end{theorem}
By revisiting many classical reductions in the context of fine-grained complexity, we transfer this lower bound to scheduling problems like \tf{\one}{}{\sumwu}.
Although lower bounds do not have the immediate practical value of an algorithm, it is clear from the results of this paper how finding new lower bounds can push research into the right direction: Our lower bound for the scheduling problem \tf{\two}{\any}{\cmax} indicated the possibility of an $\Oh{nP}$-time algorithm, but the best known algorithm (by Du and Leung~\cite{DL89}) had running time $\Oh{nP^2}$. A modification of this algorithm closes this gap.

It should be noted that all lower bounds in this paper are conditional, that is, they rely on some complexity assumption. However, all of these assumptions are reasonable in the sense that a lot of effort has been put towards refuting them. And in the unlikely case that they are indeed falsified, this would have big complexity theoretical implications.

This paper is organized as follows: We first give an overview on terminology, the related lower bounds by Abboud \etal~\cite{ABHS22} and our results in \cref{sec:preliminaries}. Then we examine scheduling problems with a single machine in \cref{sec:onemachine} and problems with two or more machines in \cref{sec:moremachines}. Finally, we give a summary as well as open problems and promising research directions in \cref{sec:conclusion}. \autoref{app:omittedproofs} holds omitted proofs, \autoref{app:strongly} includes lower bounds for strongly NP-hard problems and in \autoref{app:implications}, we explore the implications of our reductions for different objective functions.

\section{Preliminaries}
\label{sec:preliminaries}
In this section, we first introduce the \PART problem, a special case of \SSS from which many of our reductions start. Then we recall common terminology from scheduling theory and finally, we give a short overview of the recent and closely related work~\cite{ABHS22} by Abboud \etal and then briefly state our main results.

Throughout this paper, $\log$ denotes the base $2$ logarithm. Moreover, we write $[n]$ for the set of integers from 1 to $n$, i.e. $[n]:=\{1,\hdots,n\}$. If we consider a set of items or jobs $[n]$ and a subset $S\subseteq[n]$, we use $\overbar{S}=[n]\setminus S$ to denote the complement of $S$. The $\tilde{\mathcal{O}}$-notation hides poly-logarithmic factors.

\subsection{Subset Sum and Partition}
\label{subsec:ssandpartition}
In this work, we provide lower bounds for several scheduling problems; our main technique are \emph{fine-grained reductions}, which are like polynomial-time reductions, but with more care for the exact sizes and running times. With these reductions, we can transfer the (supposed) hardness of one problem to another. Most of the time, our reductions start with an instance of \SSS or \PART and construct an instance of some scheduling problem. \PART is the special case of \SSS, where the sum of all items is exactly twice the target value:
\begin{problem}{\PART}
\begin{myproblem}
\instance Items $a_{1},\ldots,a_{n}\in\mathbb{N}$.
\task Decide whether there is a subset $S\subseteq [n]$ such that $\sum_{i\in S}a_i = \sum_{i\in\bar{S}} a_i$.
\end{myproblem}
\end{problem}
In the following, we always denote the total size of all items by $A:=\sum_{i=1}^n a_i$ for \SSS and \PART. Note that we can always assume that $T\leq A$, since otherwise the target cannot be reached, even by taking all items. Moreover, in the reduction by Abboud \etal~\cite{ABHS19}, $A$ and $T$ are quite close, in particular, we can assume that $A=\text{poly}(n)T$. Hence, if we could solve \SSS in time $\Oh{2^{\delta n} A^{1-\eps}}$ for some $\eps>0$ and every $\delta>0$, this would contradict~\cref{thm:exact_subsetsum} for large enough $n$. For the details, we refer to \cref{app:omittedproofs}.
\begin{restatable}{corollary}{corsubsetsum}
\label{cor:exact_subsetsum}
For every $\eps>0$, there is a $\delta>0$ such that \SSS cannot be solved in time $\Oh{2^{\delta n} A^{1-\eps}}$, unless the SETH fails.
\end{restatable}

Using a classical reduction from \SSS to \PART that only adds two large items, we also get the following lower bound for \PART (for a detailed proof, see \cref{app:omittedproofs}):
\begin{restatable}{theorem}{partition}\label{thm:partition}
For every $\eps>0$, there is a $\delta>0$ such that \PART cannot be solved in time $\Oh{2^{\delta n} A^{1-\eps}}$, unless the SETH fails.
\end{restatable}

\subsection{Scheduling}
\label{subsec:scheduling}
In all scheduling problems we consider, we are given a number of machines and a set of $n$ jobs with processing times $p_j$, $j\in[n]$; our goal is to assign each job to (usually) one machine such that the resulting \emph{schedule} minimizes some objective.\footnote{Depending on the scheduling problem, it may also be important in which order the jobs of a machine are scheduled or whether there are gaps between the execution of consecutive jobs.} So these problems all have a similar structure: A machine model, some (optional) job characteristics and an objective function. This structure motivates the use of the three-field notation introduced by Graham \etal~\cite{GLLK1979}. Hence, we denote a scheduling problem as a triple $\alpha|\beta|\gamma$, where $\alpha$ is the machine model, $\beta$ is a list of (optional) job characteristics and $\gamma$ is the objective function. As is usual in the literature, we leave out job characteristics like due dates that are implied by the objective function, e.g. for \tf{\one}{}{\sumwu}. In this work, we mainly consider the decision variants of scheduling problems (as opposed to the optimization variants). In the decision problems, we are always given a threshold denoted by $y$ and the task is to decide whether there is a solution with value at most $y$. Note that the optimization and the decision problems are -- at least in our context -- equivalent: An algorithm for the decision problem can be used to find a solution of the optimization problem with a binary search over the possible objective values (which are always integral and bounded, here). Vice versa, an algorithm for the optimization problem can also solve the decision problem. 

In order to have a unified notation, given some job-dependent parameters $g_1,\hdots,g_n$ (e.g. processing times), we let $g_{\max}:=\max_{i\in[n]}g_i$, $g_{\min}:=\min_{i\in[n]}g_i$ and $G:=\sum_{i\in[n]}g_i$. We now briefly go over the considered machine models, job characteristics and objective functions.

As the title of this work suggests, we consider problems with a fixed number of $m$ parallel identical machines, denoted by \enquote*{$Pm$} if $m>1$ or simply \enquote*{$1$} if $m=1$. In this setting, a job has the same processing time on every machine.

In the case of \emph{rigid} and \emph{moldable jobs}, each job has a given \enquote*{$size$} and must be scheduled on that many machines or it may be scheduled on \enquote*{$any$} number of machines, respectively, needing a possibly different (usually lower) processing time when scheduled on multiple machines. Sometimes, not all jobs are available at time $0$, but instead each job $j$ arrives at its release date \enquote*{$r_j$}.\footnote{This is not to be confused with \emph{online} scheduling; we know the $r_j$'s in advance.} Similarly, jobs might have deadlines $d_j$ (i.e. due dates that may not be missed) and we must assure that \enquote*{$C_j\leq d_j$} holds for every job $j$, where $C_j$ is the \emph{completion time} of $j$. Additionally, every job $j$ might have a weight $w_j$ and we are allowed to reject (i.e., choose not to schedule) jobs of total weight at most $Q$; this constraint is denoted by \enquote*{$Rej\leq Q$}.\footnote{This is usually denoted by $Rej\leq R$, but since we will use $R$ for the sum of all release dates, we denote the total rejection weight by $Q$.}

The arguably most popular objective in scheduling is to minimize the so-called \emph{makespan} \enquote*{$C_{\max}$}, which is the largest completion time $C_j$ among all jobs $j$, i.e. the time at which all jobs are finished. In order to give the jobs different priorities, we can minimize the \emph{total (weighted) completion time} \enquote*{$\sum w_jC_j$} (\enquote*{$\sum w_jC_j$}). If there is a due date $d_j$ for each job, we might be concerned with minimizing the \emph{(weighted) number of late jobs} \enquote*{$\sum U_j$} (\enquote*{$\sum w_jU_j$}), where $U_j=1$ if $j$ is late, i.e. $C_j>d_j$ and $U_j=0$ otherwise. Similar objectives are the \emph{maximum lateness} \enquote*{$L_{\max}$} and the \emph{maximum tardiness} \enquote*{$T_{\max}$} of all jobs, where the lateness $L_j$ of job $j$ is the (uncapped) difference $C_j-d_j$ and the tardiness $T_j$ is the (capped) difference $\max\{C_j-d_j,0\}$. Another objective, the \emph{total tardiness} \enquote*{$\sum T_j$}, measures the tardiness of all jobs together and the \emph{total late work} \enquote*{$\sum V_j$} is the late work $V_j:=\min\{p_j,C_j-d_j\}$ summed over all jobs. Both objectives may also appear in combination with weights. Lastly, if release dates $r_j$ are present, we might be interested in minimizing the maximum flow time \enquote*{$F_{\max}$}, the total flow time \enquote*{$\sum F_j$} or the weighted total flow time \enquote*{$\sum w_jF_j$}. These objectives are similar to the previous ones; $F_j$, the flow time of job $j$, is defined as $F_j:=C_j-r_j$, i.e. the time that passes between $j$'s release and completion. 

Some of these objectives (and job characteristics) only appear in the appendix. It should be noted that the objective functions are partially ordered in complexity (see e.g.~\cite{LLRS89}). In \cref{app:implications}, we revisit the reductions between objective functions in the context of fine-grained complexity.

\subsection{The Scheduling Lower Bounds by Abboud \etal}
\label{subsec:relatedwork}
In their more recent work~\cite{ABHS22}, Abboud \etal show lower bounds for the problems \tf{\one}{}{\sumwu}, \tf{\one}{\rej}{\sumu}, \tf{\one}{\rej}{\tmax}, \tf{\one}{\rj, \rej}{\cmax}, \tf{\two}{}{\tmax}, \tf{\two}{}{\sumu}, \tf{\two}{\rj}{\cmax} and \tf{\two}{level\text{-}order}{\cmax}.\footnote{In \enquote*{level-order} problems, the jobs are ordered hierarchically and all jobs of one level have to be finished before jobs of higher levels can be scheduled.} From those problems, only \tf{\one}{}{\sumwu} appears in the main part of this paper, but \cref{app:implications} also contains results for \tf{\one}{\rej}{\sumu}, \tf{\one}{\rej}{\tmax}, \tf{\two}{}{\tmax} and \tf{\two}{}{\sumu}. As we will see however, the results by Abboud \etal are not directly comparable to our results.

Standard dynamic programming approaches often give running times like $\Oh{nP}$; on the other hand, it is usually possible to try out all subsets of jobs, yielding an exponential running time like $\Oh{2^{n}\text{polylog}(P)}$ (see e.g. the work by Jansen \etal~\cite{JLL16}). The intuitive way of thinking about our lower bounds is that we cannot have the best of both worlds, i.e.: \emph{\enquote*{An algorithm cannot be sub-exponential in $n$ and sub-linear in $P$ at the same time.}} 
To be more specific, most of our lower bounds have this form: For every $\eps>0$, there is a $\delta>0$ such that the problem cannot be solved in time $\Oh{2^{\delta n}P^{1-\eps}}$. 

However, note that algorithms with running time $\Ohtilde{n+P}$ or $\Ohtilde{n+p_{\max}}$ are not excluded by our bounds, as they are not sub-linear in $P$. But in a setting where $n$ and $P$ (resp. $p_{\max}$) are roughly of the same order, such algorithms would be much more efficient than the dynamic programming approaches. In particular, they would be near-linear in $n$ instead of quadratic. This is where the lower bounds from the more recent paper~\cite{ABHS22} by Abboud \etal come into play, as they have the following form: There is no $\eps>0$ such that the problem can be solved in time $\Ohtilde{n+p_{\max}n^{1-\eps}}$, unless the $\forall\exists$-SETH fails. These lower bounds can successfully exclude algorithms with an additive-type running time $\Ohtilde{n+p_{\max}}$. Algorithms with running time $\Ohtilde{n+p_{\max}n}$ may still be possible, but they would only be near-quadratic instead of near-linear in the $n\approx p_{\max}$ setting. It should be mentioned that the lower bounds by Abboud \etal~\cite{ABHS22} rely on the $\forall\exists$-SETH and as noted by them, this assumption is stronger than the SETH, even strictly stronger, if we assume the NSETH, yet another hardness assumption. For the sake of completeness, we give a detailed proof in \cref{app:omittedproofs}. Moreover, it should be noted that our lower bounds also include parameters other than $p_{\max}$, e.g. the largest due date $d_{\max}$ or the threshold for the objective value $y$.

\subsection{Our Results}
The main contribution of this work is two-fold: On the one hand, we give plenty of lower bounds for classical scheduling problems with a fixed number of machines. These lower bounds all either rely on the ETH, SETH or the $(\min,+)$-conjecture\footnote{Under the $(\min,+)$-conjecture, the $(\min,+)$-convolution problem cannot be solved in sub-quadratic time, see~\cite{CMWW19} for details.} and are shown by revisiting classical reductions in the context of fine-grained complexity, i.e., we pay much attention to the parameters of the constructed instances. On the other hand, we show how the dynamic programming algorithms for \tf{\two}{\any}{\cmax} and \tf{\three}{\any}{\cmax} by Du and Leung~\cite{DL89} can be improved. Most notably, we show the following (for the precise statements, we refer to the upcoming sections):
\begin{itemize}
    \item The algorithm by Lawler and Moore~\cite{LM69} is probably optimal for \tf{\one}{}{\sumwu} and \tf{\fixed}{}{\cmax}.
    \item The algorithm by Lee and Uzsoy~\cite{LU92} is probably optimal for \tf{\two}{}{\sumwc}.
    \item The algorithm by Zhang \etal~\cite{ZLY10} for \tf{\one}{\rej}{\sumwu}, the algorithm by Lawler~\cite{Lawler77} for \tf{\one}{}{\sumt} and the FPTAS by Gens and Levner~\cite{GL81} for \tf{\one}{}{\sumwu} are nearly optimal, but there is still some room for improvement.
    \item \tf{\two}{\any}{\cmax} can be solved in time $\Oh{nP}$ \emph{and} this is probably optimal.
    \item \tf{\three}{\any}{\cmax} can be solved in time $\Oh{nP^2}$, which greatly improves upon the $\Oh{nP^5}$-time algorithm by Du and Leung~\cite{DL89}.
\end{itemize}
Note that our SETH-based lower bounds mainly show that improvements for some pseudo-polynomial algorithms are unlikely. For problems that are strongly NP-hard, pseudo-polynomial algorithms cannot exist, unless P=NP~\cite{CPW98}. We still give lower bounds under SETH for some strongly NP-hard scheduling problems, as they exclude algorithms that are sub-exponential but super-polynomial in $n$. However, since these results are clearly not as strong as those for the weakly NP-hard problems, they can be found in \cref{app:strongly}.


\section{Problems With One Machine}\label{sec:onemachine}
In this section, we consider problems on a single machine. For these problems, the main task is to order the jobs. First, consider again the problem \tf{\one}{}{\sumwu} of minimizing the weighted number of late jobs on a single machine. With a reduction very similar to the one by Karp~\cite{Karp72}, we get the following lower bound:\footnote{It should be noted that some of the parameters in our lower bounds could be omitted, as they are overshadowed by others. For example, we can assume w.l.o.g. that $d_{\max}\leq P$ for \tf{\one}{}{\sumwu}, since we can assume a schedule to be gap-less and hence due dates larger than $P$ could be set to $P$. But having all the parameters in the lower bound makes the comparison with known upper bounds easier.}
\begin{restatable}{theorem}{onesumwu}\label{thm:1||sumw_jU_j}
For every $\eps>0$, there is a $\delta>0$ such that \tf{\one}{}{\sumwu} cannot be solved in time $\Oh{2^{\delta n} (d_{\max}+y+P+W)^{1 - \eps}}$, unless the SETH fails.
\end{restatable}
\begin{proof}
	Let $a_1, \dots, a_n$ be a \PART instance and let $T=\frac{1}{2}\sum_{i=1}^n a_i$.
	Construct an instance of \tf{\one}{}{\sumwu} by setting $p_j = w_j = a_j, d_j = T$ for each $j \in [n]$ and $y = T$.
	The idea is that the jobs corresponding to items in one of the partitions can be scheduled early (i.e. before the uniform due date $T$). For a formal proof regarding the correctness of the reduction, see \cref{app:omittedproofs}.
	
 	With this reduction, we get $N:=n$ jobs. We have $P=\sum_{i=1}^n a_i=A$ and hence $K:=d_{\max}+y+P+W=T+T+A+A=\text{poly}(n)A=n^cA$. The reduction itself takes time $\Oh{N}$. Assuming that we can solve \tf{\one}{}{\sumwu} in time $\Oh{2^{\delta N} K^{1 - \eps}}$ for some $\eps>0$ and every $\delta>0$, we could also solve \PART in time:
 	\begin{align*}
 	\Oh{N}+\Oh{2^{\delta N} K^{1 - \eps}}=\Oh{n}+\Oh{2^{\delta n} (n^cA)^{1 - \eps}} &\leq\Oh{2^{\delta n}n^cA^{1-\eps}} \\
 	&=\Oh{2^{\delta n+c\log(n)}A^{1-\eps}} \\
 	&\leq\Oh{2^{2\delta n}A^{1-\eps}}
 	\end{align*}
 	The last step holds for large enough $n$; for smaller $n$, we can solve the problem efficiently, anyway, as $n$ is then bounded by a constant. Now, to contradict \cref{thm:partition}, we can set $\eps':=\eps$ and for every $\delta'>0$, we have $\delta=\frac{\delta'}{2}>0$. So by assumption, we can solve \PART in time $\Oh{2^{2\delta n}A^{1-\eps}}=\Oh{2^{\delta' n}A^{1-\eps'}}$.
\end{proof}
Using the algorithm by Lawler and Moore~\cite{LM69}, \tf{\one}{}{\sumwu} is solvable in time $\Oh{nW}$ or $\Oh{n\min\{d_{\max},P\}}$. Our $\Oh{2^{\delta n} (d_{\max}+y+P+W)^{1 - \eps}}$-time lower bound suggests the optimality of both variants, as we cannot hope to reduce the linear dependency on $W$, $d_{\max}$ or $P$ without getting a super-polynomial dependency on $n$. As noted above, Abboud \etal~\cite{ABHS22} exclude $\Ohtilde{n+p_{\max}n^{1-\eps}}$-time algorithms; Hermelin \etal~\cite{HMS22} exclude algorithms with running time $\Ohtilde{n+w_{\max}n^{1-\eps}}$, and $\Ohtilde{n+w_{\max}^{1-\eps}n}$ and $\Ohtilde{n^{\Oh{1}}+d_{\max}^{1-\eps}}$ (all three under the stronger $\forall\exists$-SETH).

One interesting property of \tf{\one}{}{\sumwu} is that its straightforward formulation as an Integer Linear Program has a triangular structure that collapses to a single constraint when all due dates are equal (see e.g. Lenstra and Shmoys~\cite{LS20}). This shows that the problem is closely related to \KS:

\begin{problem}{\KS}
  \begin{myproblem}
    \instance Item values $v_{1},\ldots,v_{n}\in\mathbb{N}$, item sizes
    $a_{1},\ldots,a_{n}\in\mathbb{N}$, knapsack capacity $T\in\mathbb{N}$ and threshold $y$.
    \task Decide whether there is a subset $S$ of items with $\sum_{j\in S}a_{j}\leq
    T$ and $\sum_{j\in S}v_{j}\geq y$.
  \end{myproblem}
\end{problem}
Cygan \etal~\cite{CMWW19} conjectured that the \textsc{$(\min,+)$-Convolution} problem cannot be solved in sub-quadratic time (this is known as the $(\min,+)$-conjecture) and showed that this conditional lower bound transfers to \KS, excluding $\Oh{(n+T)^{2-\delta}}$ algorithms. As noted by Mucha \etal~\cite{MWW19}, these results also hold when we swap the role of sizes and values. As we can discard items with too large value $v_i$, a lower bound depending on the largest item value $v_{\max}$ directly follows from Corollary 9.6 in~\cite{MWW19}:
\begin{corollary}
  \label{cor:approx:knapsack}
  For any constant $\delta>0$, there is no $\Oh{\left(n+v_{\max}\right)^{2-\delta}}$-time exact algorithm for \KS, unless the $(\min,+)$-conjecture fails.
\end{corollary}
We show that the conditional hardness of \KS transfers to \tf{\one}{}{\sumwu}:
\begin{restatable}{theorem}{exactwu}
  \label{thm:exact:wjUj}
  For any constant $\delta>0$, the existence of an exact algorithm  
  for \tf{\one}{}{\sumwu} with running time
  $\Oh{(n+w_{\max})^{2-\delta}}$ refutes the $(\min,+)$-conjecture. 
\end{restatable}
\begin{proof}
We give a reduction from \KS to \tf{\one}{}{\sumwu}. Consider an instance $v_{1},\ldots,v_{n}$, $a_{1},\ldots,a_{n}$, $T$, $y$ of \KS. We construct jobs with $p_j=a_j$, $w_j=v_j$ and $d_j=T$ for every $j\in[n]$. The threshold is set to $y'=\sum_{j=1}^n v_j - y$. As this is also a very classical reduction, we leave the proof of correctness to the appendix.

Suppose that there is an $\Oh{(n+w_{\max})^{2-\delta}}$-time algorithm for \tf{\one}{}{\sumwu}. Since $w_{\max}=v_{\max}$ in the reduction and the reduction takes time $\Oh{n}$, we could then solve \KS in time $\Oh{n}+\Oh{(n+w_{\max})^{2-\delta}}=\Oh{\left(n+v_{\max}\right)^{2-\delta}}$,
which is a contradiction to~\cref{cor:approx:knapsack}, unless the $(\min,+)$-conjecture fails. 
\end{proof}

Lower bounds such as this one also imply lower bounds for approximation schemes, as setting the accuracy parameter $\eps$ small enough yields an exact solution. The above result implies the following (see~\cref{app:omittedproofs} for the proof):
\begin{restatable}{corollary}{approxsumwu} \label{cor:approx:1||sumwu}
  For any constant $\delta>0$, the existence of an $\Oh{(n+\frac{1}{2n\eps})^{2-\delta}}$-time approximation scheme for the optimization version of \tf{\one}{}{\sumwu} refutes the $(\min,+)$-conjecture.
\end{restatable}
As the currently fastest FPTAS by Gens and Levner~\cite{GL81} runs in time $\Oh{n^{2}(\log(n)+\frac{1}{\eps})}$, there is still a small gap. This relation between exact and approximation algorithms might also be an interesting subject of further investigation, as many other scheduling problems admit approximation schemes and exact lower bounds.

We wish to mention two other results, the proofs of which can also be found in \cref{app:omittedproofs}. The first result concerns \tf{\one}{}{\sumt}:
\begin{restatable}{theorem}{totaltardiness}\label{thm:1||sumT_j}
For every $\eps>0$, there is a $\delta>0$ such that \tf{\one}{}{\sumt} cannot be solved in time $\Oh{2^{\delta n} P^{1-\eps}}$, unless the SETH fails.
\end{restatable}
There is an $\Oh{n^4P}$-time algorithm by Lawler~\cite{Lawler77} and while we can derive no statement about the exponent of $n$, our lower bound suggests that an improvement of the linear factor $P$ is unlikely without getting a super-polynomial dependency on $n$. We have a similar situation for the problem \tf{\one}{\rej}{\cmax}:
\begin{restatable}{theorem}{reject}\label{thm:1|Rej|C_{max}}
For every $\eps>0$, there is a $\delta>0$ such that \tf{\one}{\rej}{\cmax} cannot be solved in time $\Oh{2^{\delta n} (y+P+Q+W)^{1-\eps}}$, unless the SETH fails.
\end{restatable}
The lower bound can also be shown to hold for \tf{\one}{\rej}{\sumwu} (see \cref{app:implications}) and this problem can be solved in time $\Oh{nQP}$ with the algorithm by Zhang \etal~\cite{ZLY10}. This almost matches our lower bound: An algorithm with running time $\Oh{n(Q+P)}$ might still be possible, for example.

\section{Problems With Multiple Machines}
\label{sec:moremachines}
We now turn our attention to problems on two or more machines.
For \textbf{standard jobs}, a straightforward reduction from \PART yields the following result (for a formal proof, see \cref{app:omittedproofs}):
\begin{restatable}{theorem}{twocmax}\label{thm||C_{max}}
	For every $\eps>0$, there is a $\delta>0$ such that \tf{\two}{}{\cmax} cannot be solved in time $\Oh{2^{\delta n} (y+P)^{1-\eps}}$, unless the SETH fails.
\end{restatable}
This lower bound also applies to the harder objectives (e.g. $T_{\max}$) and in particular to \tf{\two}{}{\sumwu} (see \cref{app:implications}); the dynamic program by Lawler and Moore~\cite{LM69} (which is also sometimes attributed to Rothkopf~\cite{Rothkopf66}) solves most common objectives like $C_{\max}$ and $T_{\max}$ in time $\Oh{ny}$ but needs $\Oh{ny^2}$ for \tf{\two}{}{\sumwu} (see~\cite{LS20}, in particular exercise 8.10). So the gap is likely closed in the $C_{\max}$, $T_{\max}$, $\hdots$-cases, but there is still a factor-$y$-gap for the $\sum w_jU_j$-objective.

In general, the dynamic program by Lawler and Moore~\cite{LM69} solves \tf{\fixed}{}{\cmax} in a running time of $\Oh{nmy^{m-1}}\leq\Oh{nmP^{m-1}}$. Our matching lower bound for $m=2$ gives rise to the question whether the running time is optimal for general $m>1$. In \cref{app:omittedproofs}, we prove the following result:
\begin{restatable}{theorem}{fixedcmax}\label{thm:fixed}
  There is no $\Oh{nmP^{\oh{\frac{m}{\log^2(m)}}}}$-time algorithm for \tf{\fixed}{}{\cmax}, unless the ETH fails.
\end{restatable}
So the algorithm by Lawler and Moore~\cite{LM69} is indeed almost optimal, as we can at best hope to shave off logarithmic factors in the exponent (assuming the weaker assumption ETH). Since the algorithm not only works for $C_{\max}$, one might ask whether we can find similar lower bounds for other objectives as well. For most common objective functions, we answer this question positively in \cref{app:implications}, but it remains open for $\sum w_jC_j$. Note that the unweighted \tf{\fixed}{}{\sumc} is polynomial-time solvable~\cite{BCS74}.

An alternative dynamic program by Lee and Uzsoy~\cite{LU92} solves $Pm||\sum w_jC_j$ in time $\Oh{mnW^{m-1}}$. In order to get a matching lower bound (i.e. one that depends on the weights) for $m=2$, we examine another classical reduction:
\begin{restatable}{theorem}{ptwosumwc}\label{thm||sumw_jC_j}
	For every $\eps>0$, there is a $\delta>0$ such that \tf{\two}{}{\sumwc} cannot be solved in time $\Oh{2^{\delta n} (\sqrt{y}+P+W)^{1-\eps}}$, unless the SETH fails.
\end{restatable}

\begin{proof}
	We show that the lower bound for \PART can be transferred to \tf{\two}{}{\sumwc} using the reduction by Lenstra \etal~\cite{LRB77} and Bruno \etal~\cite{BCS74}.
	
	Given a \PART instance $a_1, \dots, a_n$, we construct a \tf{\two}{}{\sumwc} instance in the following way: Define $p_j = w_j = a_j$ for all $j\in[n]$ and set the limit $y = \sum_{1\leq i \leq j \leq n} a_j a_i - \frac{1}{4}A^2$. Of course, the idea of the reduction is that the limit $y$ forces the jobs to be equally distributed among the two machines (regarding the processing time). We formally prove the correctness of the reduction in the appendix.
	
	Assume that there is an algorithm that solves an instance of \tf{\two}{}{\sumwc} in time $\Oh{2^{\delta N}K^{1-\eps}}$ for some $\eps>0$ and every $\delta>0$, where $N:=n$ and $K:=\sqrt{y}+P+W$. By the choice of $y$, we can see that
    \[y=\sum_{1\leq i \leq j \leq n} a_j a_i - \frac{1}{4}A^2\leq\left(\sum_{j\in[n]}a_j\right)^2-\frac{1}{4}A^2=\frac{3}{4}A^2=\Oh{A^2}.\]
    Since $w_j=p_j=a_j$, we also have $P=W=A$. Hence, we have $K=\sqrt{y}+P+W=\Oh{A+A+A}=\Oh{A}$ and an algorithm with running time
    \[\Oh{2^{\delta N}K^{1-\eps}}=\Oh{2^{\delta n}\Oh{A}^{1-\eps}}=\Oh{2^{\delta n}c^{1-\eps}A^{1-\eps}}=\Oh{2^{\delta n}A^{1-\eps}}\]
    would contradict the lower bound for \PART from \cref{thm:partition}. Here, $c$ covers the constants in the $\mathcal{O}$-term and the running time $\Oh{N}$ of the reduction vanishes.
\end{proof}
So the $\Oh{nW}$-time algorithm by Lee and Uzsoy~\cite{LU92} is probably optimal for \tf{\two}{}{\sumwc}, as we cannot hope to reduce the linear dependency on $W$ without getting a super-polynomial dependency on $n$.

We briefly turn our attention towards \textbf{rigid jobs}. Clearly, \tf{\two}{\size}{\cmax} is a generalization of \tf{\two}{}{\cmax} (the latter problem simply does not have two-machine jobs), so we get the following lower bound (for a formal proof, see \cref{app:omittedproofs}):
\begin{restatable}{theorem}{twosize}\label{thm:P2|size|C_{max}}
	For every $\eps>0$, there is a $\delta>0$ such that \tf{\two}{\size}{\cmax} cannot be solved in time $\Oh{2^{\delta n} (y+P)^{1-\eps}}$, unless the SETH fails.
\end{restatable}
Similarly, the algorithm by Lawler and Moore~\cite{LM69} can be used to find a feasible schedule for the one-machine jobs and the two-machine jobs can be scheduled at the beginning. This gives an $\Oh{ny}$-time algorithm for \tf{\two}{\size}{\cmax}, and the linear dependency on $y$ cannot be improved without getting a super-polynomial dependency on $n$, unless the SETH fails. 
For other objectives, the problem quickly becomes more difficult: Already \tf{\two}{\size}{\lmax} is strongly NP-hard, as well as \tf{\two}{\size}{\sumwc} (for both results, see Lee and Cai~\cite{LC99}). It is still open whether the unweighted version \tf{\two}{\size}{\sumc} is also strongly NP-hard or whether there is a pseudo-polynomial algorithm; this question has already been asked by Lee and Cai~\cite{LC99}, more than 20 years ago.

It is not hard to see that the hardness of \tf{\two}{}{\cmax} also transfers to \textbf{moldable jobs} (i.e. \tf{\two}{\any}{\cmax}); we simply create an instance where it does not make sense to schedule any of the jobs on two machines (for a formal proof, see \cref{app:omittedproofs}):
\begin{restatable}{theorem}{twoany}\label{thm:P2|any|C_{max}}
	For every $\eps>0$, there is a $\delta>0$ such that \tf{\two}{\any}{\cmax} cannot be solved in time $\Oh{2^{\delta n} (y+P)^{1-\eps}}$, unless the SETH fails.
\end{restatable}
The problems \tf{\two}{\any}{\cmax} and \tf{\three}{\any}{\cmax} can be solved via dynamic programming, as shown by Du and Leung~\cite{DL89} (a nice summary is given in the book by Drozdowski~\cite{Drozdowski09}). We show that these programs can be improved to match our new lower bound for the two-machine case: 
\begin{theorem}\label{thm:dp_two_machines}
  The problem \tf{\two}{\any}{\cmax} can be solved in time $\Oh{nP}$ via dynamic programming.
\end{theorem}
\begin{proof}
Assume that we are given processing times $p_j(k)$, indicating how long it takes to run job $j$ on $k$ machines. The main difficulty is to decide whether a job is to be processed on one or on two machines. Our dynamic program fills out a table $F(j,t)$ for every $j\in[n]$ and $t\in[y]$, where the entry $F(j,t)$ is the minimum load we can achieve on machine~$2$, while we schedule all the jobs in $[j]$ and machine $1$ has load $t$. To fill the table, we use the following recurrence formula:
\begin{align*}
    F(j,t)= \min\begin{cases}
        F(j-1,t-p_j(1)) \\
        F(j-1,t)+p_j(1) \\
        F(j-1,t-p_j(2))+p_j(2)
    \end{cases}
\end{align*}
Intuitively speaking, job $j$ is executed on machine~$1$ in the first case, on machine $2$ in the second case and on both machines in the third case. The initial entries of the table are $F(0,0)=0$ and $F(0,t)=\infty$ for every $t\in[y]$.

There are $ny\leq n\sum_{j=1}^n\max\{p_j(1),p_j(2)\}=\Oh{nP}$ entries we have to compute.\footnote{The precise definition of $P$ in this context does not matter for the running time in $\mathcal{O}$-notation; we can either add both $p_j(1)$ and $p_j(2)$ to the sum or just the larger of the two.} Then, we can check for every $t\in[y]$ whether $F(n,t)\leq y$. If we find such an entry, this directly corresponds to a schedule with makespan at most $y$, so we can accept. Otherwise, there is no such schedule and we can reject. The actual schedule can be obtained by traversing backwards through the table; alternatively, we can store the important bits of information while filling the table (this works exactly like in the standard knapsack algorithm). Note that we might have to reorder the jobs such that the jobs executed on two machines are run in parallel. But it can be easily seen that all two-machine jobs can be executed at the begin of the schedule. Computing the solution and reordering does not change the running time in $\mathcal{O}$-notation, so we get an $\Oh{nP}$ algorithm.
\end{proof}
As \cref{thm:P2|any|C_{max}} shows, improving the dependency on $P$ to sub-linear is only possible if we get a super-polynomial dependency on $n$, unless the SETH fails. In a similar way, one can also improve the dynamic program for three machines (the proof is given in~\cref{app:omittedproofs}):
\begin{restatable}{theorem}{dynprog}\label{thm:dp_three_machines}
  The problem \tf{\three}{\any}{\cmax} can be solved in time $\Oh{n^2P}$ via dynamic programming.
\end{restatable}
This improves upon the $\Oh{nP^5}$-algorithm by Du and Leung~\cite{DL89}. Even though the same approach could be applied to an arbitrary number of machines $m$ in time $\Oh{nmP^{m-1}}$, the strong NP-hardness of $Pm|any|C_{\max}$ for $m\geq 4$ shows that the information on which machine each job is scheduled is not enough to directly construct an optimal schedule in those cases, unless P=NP (see Henning \etal~\cite{HJRS19} as well as Du and Leung~\cite{DL89}).

\section{Conclusion}\label{sec:conclusion}
In this work, we examined the complexity of scheduling problems with a fixed number of machines. Our conditional lower bounds indicate the optimality of multiple well-known classical algorithms. For the problems \tf{\two}{\any}{\cmax} and \tf{\three}{\any}{\cmax}, we managed to improve the currently best known algorithm, closing the gap for two machines. 

As we have seen in the example of \tf{\one}{}{\sumwu}, lower bounds for exact algorithms can be quite easily used to obtain lower bounds for approximation schemes. We strongly believe that the same technique can be used for other problems, either to show tightness results or to indicate room for improvement.

For exact algorithms, there is a number of open problems motivated by our results: First of all, there is still a gap between our lower bound and the algorithm by Lawler and Moore~\cite{LM69}. So an interesting question is where the \enquote*{true} complexity lies between $m-1$ and $\oh{\frac{m}{\log^2(m)}}$ in the exponent. 
Zhang \etal give an $\Oh{n(r_{\max}+P)}$-time algorithm for \tf{\one}{\rj, \rej}{\cmax} in their work~\cite{ZLY10}. Since $r_{\max}+P\geq y$ w.l.o.g., it would be interesting to find an $\Oh{2^{\delta n} (r_{\max}+P)^{1 - \eps}}$ or $\Oh{2^{\delta n} y^{1 - \eps}}$ lower bound for this problem. As noted by Lenstra and Shmoys~\cite{LS20}, the algorithm by Lawler and Moore~\cite{LM69} cannot be improved to $\Oh{mny^{m-1}}$ for the objective $\sum w_jU_j$. So this algorithm would be quadratic in $y$ for two machines, while our lower bound excludes anything better than linear (and still polynomial in $n$). Hence, it would be interesting to see whether there is a different algorithm with running time $\Oh{ny}$. Similarly, there is an algorithm for \tf{\one}{\rej}{\sumwu} with running time $\Oh{nQP}$\cite{ZLY10}, while our lower bound suggests that an $\Oh{n(Q+P)}$-time algorithm could be possible. 
 
On another note, it would be interesting to extend the sub-quadratic equivalences by Cygan \etal~\cite{CMWW19} and Klein~\cite{Klein21} to scheduling problems. Finally, the question by Lee and Cai~\cite{LC99} whether \tf{\two}{\size}{\sumc} is strongly NP-hard or not is still open since 1999.

\bibliography{bib}

\appendix

\section{Omitted Proofs}\label{app:omittedproofs}
In this section, we give the proofs that were omitted from the main part. These are mostly reductions and the dynamic program for \tf{\three}{\size}{\cmax}, but we also show that the $\forall\exists$-SETH is probably a strictly stronger assumption than SETH. 

\subsection{SETH and \texorpdfstring{$\forall\exists$}{Forall-Exists}-SETH}
Before we prove \cref{prop:forallexists}, we restate the SETH by Impagliazzo and Paturi~\cite{IP99} and the $\forall\exists$-SETH by Abboud \etal~\cite{ABHS22}:
\begin{conjecture}[Strong Exponential Time Hypothesis~\cite{IP99}]
  For every $\eps>0$, there is some $k\geq3$ such that \textsc{$k$-Sat} cannot be solved in time $\Oh{2^{(1-\varepsilon)n}}$.
\end{conjecture}

\begin{conjecture}[$\forall\exists$ Strong Exponential Time Hypothesis~\cite{ABHS22}]
  For every $\alpha\in(0,1)$, $\eps>0$ there is some $k\geq3$ such that the problem of deciding whether
  \[\forall x_1,\hdots,x_{\lceil \alpha n\rceil}\exists x_{\lceil \alpha n\rceil +1},\hdots,x_n\,:\,\phi(x_1,\hdots,x_n)=\text{true}\]
  cannot be solved in time $\Oh{2^{(1-\varepsilon)n}}$ for any $n$-variable formula $\phi$ in conjunctive normal form with $k$ variables per clause.
\end{conjecture}

We show the \enquote*{strictly stronger} part of the claim using the Non-Deterministic Strong Exponential Time Hypothesis (NSETH):
\begin{conjecture}[Non-Deterministic Strong Exponential Time Hypothesis~\cite{CGIRMPS16}]
  For every $\eps>0$, there exists a $k$ such that there is no non-deterministic algorithm solving the complement of \textsc{$k$-Sat} in time $\Oh{2^{(1-\varepsilon)n}}$.
\end{conjecture}
This conjecture is particularly useful for proving non-reducibility results: If there are non-deterministic algorithms for a problem $A$ and its complement $\overbar{A}$, both with running time bounded by $T$, then we cannot prove a SETH-based lower bound for $A$ that is higher than $T$, assuming NSETH (see Corollary 2 in~\cite{CGIRMPS16}). As noted by Abboud \etal~\cite{ABHS22}, under NSETH, $\forall\exists$-SETH is a stronger assumption than SETH. We provide a more detailed proof, here.

\begin{proposition}\label{prop:forallexists}
  $\forall\exists$-SETH implies SETH. But SETH does not imply $\forall\exists$-SETH, unless NSETH fails.
\end{proposition}
\begin{proof}
We prove the first part indirectly by showing how a faster-than-SETH algorithm for \textsc{$k$-Sat} would imply a faster-than-$\forall\exists$-SETH algorithm for \textsc{$\forall\exists$-$k$-SAT}. The second part is then shown by providing non-deterministic algorithms for \textsc{$\forall\exists$-$k$-SAT} and $\overbar{\textsc{$\forall\exists$-$k$-SAT}}$, which under NSETH rules out a corresponding reduction from \textsc{$k$-Sat}.

Now, assume that SETH does not hold (i.e. there is an $\eps>0$ such that \textsc{$k$-Sat} can be solved in time $\Oh{2^{(1-\eps)n}}$ for every $k$) and consider an instance of \textsc{$\forall\exists$-$k$-SAT}, consisting of an $\alpha\in(0,1)$, a number $k\in\mathbb{N}$ and a $k$-CNF formula $\phi(x_1,\hdots,x_n)$ depending on $n=n_1+n_2$ variables, where $n_1=\lceil\alpha n\rceil\leq \alpha n + 1$ and $n_2=n-n_1\leq (1-\alpha)n$. 

Given the formula $\phi$, we go through all $2^{n_1}$ assignments of the $n_1$ $\forall$-quantified variables and for each of them we fix the corresponding variables in $\phi$, i.e. for every appearance of a variable $x_j$ in some clause, we either remove that clause (since the clause is already satisfied) or we remove the literal from the clause (since the literal is false). Going through all the clauses takes time $\Oh{kn^k}$, as we have $\Oh{n^k}$ clauses and (at most) $k$ literals per clause. By fixing the $n_1$ variables, we get a formula with $n_2$ variables. Now, as we assumed SETH to be false, we can solve this formula in time $\Oh{2^{(1-\eps)n_2}}$ for some $\eps>0$ and large enough (but constant) $k$.

In total, we need the following running time to solve the \textsc{$\forall\exists$-$k$-SAT}-problem:
\begin{align*}
    \Oh{2^{n_1}(\Oh{kn^k}+2^{(1-\eps)n_2}}) &\leq \Oh{2^{n_1+k\log(kn)+(1-\eps)n_2}} \\
    &= \Oh{2^{n_1+k\log(kn)+n_2-\eps n_2}} \\
    &= \Oh{2^{n+k\log(kn)-\eps n_2}} \\
    &= \Oh{2^{n+k\log(kn)-\eps (n-n_1)}} \\
    &= \Oh{2^{n+k\log(kn)-\eps n+\eps n_1)}} \\
    &\leq \Oh{2^{n+k\log(kn)-\eps n+\eps (\alpha n + 1)}} \\
    &= \Oh{2^{n+k\log(kn)-\eps n+\eps\alpha n + \eps}} \\
    &= \Oh{2^{n(1+\frac{k\log(kn)+\eps}{n}-\eps+\eps\alpha)}} \\
    &= \Oh{2^{n(1-(-\frac{k\log(kn)+\eps}{n}+\eps-\eps\alpha))}}
\end{align*}
So if we can assure that $\eps':=-\frac{k\log(kn)+\eps}{n}+\eps-\eps\alpha>0$, we have contradicted the $\forall\exists$-SETH. This is exactly the case if $\eps-\eps\alpha>\frac{k\log(kn)+\eps}{n}$. Since $\alpha\in(0,1)$ and $\eps>0$, we have $\eps>\eps\alpha$ and hence $\eps-\eps\alpha>0$. And as $k$ and $\eps$ are constant, the inequality $n>\frac{k\log(kn)+\eps}{\eps-\eps\alpha}$ holds for large enough $n$. If it does not hold, $n$ has to be bounded by some constant and we can solve the \textsc{$\forall\exists$-$k$-SAT} problem efficiently, anyway. Hence, if SETH fails, $\forall\exists$-SETH fails as well and we have shown the first part of the claim.

For the second part, consider an instance of \textsc{$\forall\exists$-$k$-SAT}, consisting of $\alpha\in(0,1)$, $k\in\mathbb{N}$ and $\phi(x_1,\hdots,x_n)$ depending on $n=n_1+n_2$ variables, where $n_1=\lceil\alpha n\rceil\leq \alpha n +1$ and $n_2=n-n_1\leq (1-\alpha)n$.

To define a non-deterministic algorithm for \textsc{$\forall\exists$-$k$-SAT}, we proceed similar to the above reduction: We try out all $2^{n_1}$ assignments for the $\forall$-quantified variables and fix the corresponding variables in the given formula in time $\Oh{kn^k}$, as we need to go through all clauses and literals. We proceed to guess a satisfying assignment of the remaining $n_2$ variables in time $\Oh{n_2}$. In total, using the bounds for $n_1, n_2$, this yields a non-deterministic algorithm for \textsc{$\forall\exists$-$k$-SAT} with running time

\begin{align*}
    2^{n_1}(\Oh{kn^k}+\Oh{n_2}) &\leq \Oh{2^{n_1+k\log(nk)+\log(n_2)}}\\
    &\leq \Oh{2^{\alpha n + 1 +k\log(nk)+\log((1-\alpha)n)}} \\
    &= \Oh{2^{n(\alpha + \frac{k\log(nk)+1+\log((1-\alpha)n)}{n}}} \\
    &= \Oh{2^{n(1-(1-\alpha - \frac{k\log(nk)+1+\log((1-\alpha)n)}{n}}}
\end{align*}
and setting $\eps:=1-\alpha - \frac{k\log(nk)+1+\log((1-\alpha)n)}{n}$ gives us the desired running time of $\Oh{2^{(1-\eps)n}}$, but we have to again assure that $\eps>0$. This holds if and only if $1-\alpha>\frac{k\log(nk)+1+\log((1-\alpha)n)}{n}$. But again, $1-\alpha>0$ and if the inequality $n>\frac{k\log(nk)+1+\log((1-\alpha)n)}{1-\alpha}$ does not hold, $n$ is bounded by some constant and the problem can be solved efficiently.

In a non-deterministic algorithm for $\overbar{\textsc{$\forall\exists$-$k$-SAT}}$, we need to decide whether there exists an assignment for the first $n_1$ variables such that for every assignment of the remaining $n_2$ variables, the formula $\phi$ evaluates to false. Up to changes in the order of the steps, the algorithm works almost identical to the above one: We first guess a feasible assignment for the $n_1$ variables in time $\Oh{n_1}$ and then we try out all $2^{n_2}$ assignments for the $n_2$ variables and evaluate the resulting formulas in time $\Oh{kn^k}$ by going through each clause and literal. So we get the following running time:
\begin{align*}
    \Oh{n_1} + 2^{n_2}\Oh{kn^k} &\leq \Oh{2^{\log(n_1)+n_2+k\log(nk)}}\\
    &\leq \Oh{2^{\log(\alpha n +1)+(1-\alpha)n+k\log(nk)}} \\
    &= \Oh{2^{n(\frac{\log(\alpha n +1)+k\log(nk)}{n}+(1-\alpha)}} \\
    &= \Oh{2^{n(1-(-\frac{\log(\alpha n +1)+k\log(nk)}{n}+\alpha))}}
\end{align*}
Again, setting $\eps:=-\frac{\log(\alpha n +1)+k\log(nk)}{n}+\alpha$ yields a running time of $\Oh{2^{(1-\eps)n}}$. We get $\eps>0$ if and only if $\alpha>\frac{\log(\alpha n +1)+k\log(nk)}{n}$, which holds for large enough $n$. If the inequality does not hold, we can also solve the problem efficiently, as $n$ is then bounded by a constant.

Hence, there is an $\eps>0$ such that both \textsc{$\forall\exists$-$k$-SAT} and $\overbar{\textsc{$\forall\exists$-$k$-SAT}}$ can be solved in time $\Oh{2^{(1-\eps)n}}$ (we just take the smaller of the two), which implies that there is no $\Oh{2^{(1-\eps)n}}$ lower bound via SETH, unless NSETH fails (see Corollary 2 in~\cite{CGIRMPS16}).
\end{proof}

\subsection{Weakly NP-hard Problems}
We now give the proofs that did not make it into the main part of this paper. First of all, we prove \cref{cor:exact_subsetsum}:
\corsubsetsum*
\begin{proof}
It is important that $A=\text{poly}(n)T=n^cT$ for some constant $c$ in the reduction by Abboud \etal~\cite{ABHS19}. Assume that there is an $\eps>0$ such that for every $\delta>0$, \SSS can be solved in time $\Oh{2^{\delta n}A^{1-\eps}}$. We have
\begin{align*}
    \Oh{2^{\delta n}A^{1-\eps}} = \Oh{2^{\delta n}(n^cT)^{1-\eps}} \leq \Oh{2^{\delta n}n^cT^{1-\eps}}
    &= \Oh{2^{\delta n+c\log(n)}T^{1-\eps}} \\
    &\leq \Oh{2^{2\delta n}T^{1-\eps}} \\
    &= \Oh{2^{\delta' n}T^{1-\eps}}
\end{align*}
if we set $\delta':=2\delta$ and assume that $n$ is large enough so that $\delta n \geq c\log(n)$. Otherwise, $n$ is bounded by a constant depending on $\delta$ and $c$. Note that we assume that we can solve \SSS \textit{for any} $\delta>0$. Hence, for every $\delta'$, we can find a $\delta=\frac{\delta'}{2}$ and get a contradiction to \cref{thm:exact_subsetsum}.
\end{proof}

To avoid repetitions, we now show a useful lemma that encapsulates the technical parts in the computations of our lower bounds:
\begin{lemma}\label{lem:reductions}
Suppose there is an $\Oh{\text{poly}(n)}$-time reduction from \SSS (or \PART) with $n$ items and $\sum_{i=1}^na_i=A$ to some scheduling problem $\alpha|\beta|\gamma$ with $N=\Oh{n}$ jobs and parameter $K=\Oh{\text{poly}(n)A}$. Then for every $\eps>0$, there exists a $\delta>0$ such that $\alpha|\beta|\gamma$ cannot be solved in time $\Oh{2^{\delta N}K^{1-\eps}}$, unless the SETH fails.
\end{lemma}
\begin{proof}
  For the sake of contradiction, assume that there exists an $\eps>0$ such that for every $\delta>0$, $\alpha|\beta|\gamma$ can be solved in time $\Oh{2^{\delta N}K^{1-\eps}}$. Now, consider an instance of \SSS (or \PART) with $n$ items and $\sum_{i=1}^na_i=A$. Using the reduction, we construct an instance of $\alpha|\beta|\gamma$ with $N=\Oh{n}=c_1n$ jobs and parameter $K=\Oh{\text{poly}(n)A}=c_2n^{c_3}A$ in time $\Oh{\text{poly}(n)}=n^{c_4}$. 
  
  In order to contradict the lower bound for \SSS (or \PART), we set $\eps':=\eps$ and consider some arbitrary but fixed $\delta'>0$. Since we can -- by assumption -- solve $\alpha|\beta|\gamma$ in time $\Oh{2^{\delta N}K^{1-\eps}}$ for \emph{every} $\delta>0$, we can also do so for $\delta=(\delta'-\frac{(c_3+c_4)\log(n)}{n})\frac{n}{c_1 n}$, as long as this is larger than $0$. For this, we need that $n$ is large enough so we get:
  \begin{align*}
      \left(\delta'-\frac{(c_3+c_4)\log(n)}{n}\right)\frac{n}{c_1 n} &> 0 \\
      \delta'-\frac{(c_3+c_4)\log(n)}{n} &> 0 \\
      n &> \frac{(c_3+c_4)\log(n)}{\delta'}
  \end{align*}
  Note that for smaller $n$, the inequality $n\leq \frac{(c_3+c_4)\log(n)}{\delta'}$ means that $n$ has to be bounded by some function in $\delta'$ and hence (since $\delta'$ is fixed), we could solve \SSS (or \PART) in polynomial time. So let us now assume that $n$ is large enough so that $\delta>0$ and we can use the supposed algorithm for $\alpha|\beta|\gamma$. Using the reduction and this algorithm, we can then solve the \SSS (or \PART) instance in time:
  \begin{align*}
      n^{c_4}+\Oh{2^{\delta N}K^{1-\eps}} &\leq \Oh{2^{\delta c_1 n}(c_2 n^{c_3+c_4}A)^{1-\eps}} \\
      &\leq \Oh{2^{\delta c_1 n}c_2 n^{c_3+c_4}A^{1-\eps}} \\
      &= \Oh{2^{\delta c_1 n}2^{(c_3+c_4)\log(n)}A^{1-\eps}} \\
      &= \Oh{2^{\delta c_1 n+(c_3+c_4)\log(n)}A^{1-\eps}} \\
      &= \Oh{2^{\left(\delta'-\frac{(c_3+c_4)\log(n)}{n}\right)\frac{n}{c_1 n} c_1 n+(c_3+c_4)\log(n)}A^{1-\eps'}} \\
      &= \Oh{2^{\delta'n}A^{1-\eps'}} 
  \end{align*}
  So there exists a fixed $\eps'>0$ such that for every fixed $\delta'>0$, we can solve \SSS (or \PART) in time $\Oh{2^{\delta'n}A^{1-\eps'}}$, which contradicts the corresponding lower bound under SETH and concludes the proof.
  
  Analogously, we can also get a lower bound if $K=\Oh{\text{poly}(n)T}$. We then get a contradiction to \cref{thm:exact_subsetsum}, instead.
\end{proof}

\partition*
\begin{proof}
We use a simple reduction from \SSS to \PART (a similar reduction from \KS to \PART has been given by Karp~\cite{Karp72}.
Consider a \SSS instance with items $a_1,\hdots,a_n$ and target $T$. We construct a \PART instance by copying all items $a_i'=a_i$ for all $i\in[n]$ and then adding the two items $a'_{n+1}=T+1$ and $a'_{n+2}=A+1 -T$ to the instance. Let $N=n+2$.

Given a solution $S$ of the \SSS instance, we get the partitions $S\cup\{a'_{n+2}\}$ and $\overbar{S}\cup\{a'_{n+1}\}$, which both sum up to $A+1$. For the other direction, note that the sum of all items is equal to $2A+2$ and hence the items $a'_{n+1}$ and $a'_{n+2}$ cannot be in the same partition, as they sum up to $T+1+A+1-T=A+2$. So given a solution $S\cup\overbar{S}$ of the \PART instance, assume w.l.o.g. that $a'_{n+1}$ is in $\overbar{S}$ and $a'_{n+2}$ is in $S$. Then in order for the items in $S$ to have a total sum of $A+1$, the other items in $S$ need to have a total sum that is exactly $T$. Hence, those items give us a solution of the original \SSS instance.

With $K:=\sum_{i\in[N]} a_i' = A+(T+1)+(A+1-T)=\Oh{A}$ as parameter and $N:=n+2$ jobs, \cref{lem:reductions} yields the claim, since the reduction takes time $\Oh{n}$.\footnote{Note that we only use the \SSS-part of \cref{lem:reductions}. This way, we do not actually use a lemma to prove a result that is used by the lemma itself.}
\end{proof}

\onesumwu*
\begin{proof}
    We only show the correctness of the reduction, here. The implication regarding the lower bound has already been shown above. Let $a_1, \dots, a_n$ be a \PART instance and let $T=\frac{1}{2}\sum_{i=1}^n a_i$.
	We construct an instance of \tf{\one}{}{\sumwu} by setting $p_j = w_j = a_j, d_j = T$ for each $j \in [n]$ and $y = T$.
	Remember that the idea was that the jobs corresponding to items in one of the partitions can be scheduled early (i.e. before the uniform due date $T$).
	
	Formally, assume that there is a solution $S$ of the given \PART instance. We schedule the jobs corresponding to items in $S$ first, in any order; after that, we schedule the rest of the jobs (also in any order). Now the items in $S$ sum up to $T$, so they finish exactly at $T$ and are all early. The other jobs are all late and have total weight $T=y$.
	
	For the other direction, indirectly assume that there is no solution for the \PART instance and consider any optimal schedule for the constructed instance. Without loss of generality, there are no gaps in the schedule, as they can only increase the weighted number of late jobs. Since there is no subset of items with total size exactly $T$, there is also no set of jobs with total processing time exactly $T$. Let $S$ be the set of jobs that are scheduled early and note that $\sum_{j\in S}p_j<T$. Now the schedule has total weighted number of late jobs
	\[\sum_{j=1}^n w_jU_j=\sum_{j=1}^nw_j - \sum_{j\in S}w_j=\sum_{j=1}^np_j - \sum_{j\in S}p_j>\sum_{j=1}^np_j - T=T=y,\]
	which means that an optimal schedule has value larger than $y$ and hence the instance is negative.
\end{proof}

\exactwu*
\begin{proof}
The lower bound has already been shown, so we only prove the correctness of the reduction, here. Remember that given an instance $v_{1},\ldots,v_{n}$, $a_{1},\ldots,a_{n}$, $T$, $y$ of \KS, we construct jobs with $p_j=a_j$, $w_j=v_j$ and $d_j=T$ for every $j\in[n]$. The threshold is set to $y'=\sum_{j=1}^n v_j - y$.

Let $S\subseteq[n]$ be a solution of a given \KS instance, i.e. $\sum_{j\in S}a_{j}\leq T$ and $\sum_{j\in S}v_{j}\geq y$. In the constructed \tf{\one}{}{\sumwu} instance, we schedule the jobs corresponding to items in $S$ first (in any order) and afterwards the jobs in $\overbar{S}$ (also in any order). Now, since $\sum_{j\in S}a_{j}\leq T$ and $p_j=a_j$ for every job $j$, we can see that all jobs corresponding to items in $S$ are early. The weighted number of late jobs in the schedule is therefore at most:
\[\sum_{j\in \overbar{S}} w_j=\sum_{j\in \overbar{S}} v_j=\sum_{j=1}^n v_j - \sum_{j\in S} v_j \leq \sum_{j=1}^n v_j - y=y'\]
Hence, the constructed instance of \tf{\one}{}{\sumwu} is positive.

Now, consider a solution of a constructed \tf{\one}{}{\sumwu} instance, i.e. a schedule with weighted number of late jobs at most $y'$. Let $S\subseteq[n]$ be the set of jobs that are scheduled early. Now, $S$ is a solution of the original \KS instance, since $\sum_{j\in S} a_j=\sum_{j\in S} p_j\leq T$ and $\sum_{j\in S} v_j = \sum_{j=1}^n v_j - \sum_{j\in \overbar{S}} v_j=\sum_{j=1}^n v_j - \sum_{j\in \overbar{S}} w_j\geq \sum_{j=1}^n v_j - y'=\sum_{j=1}^n v_j - \left(\sum_{j=1}^n v_j - y\right)=y$.
So the original \KS instance is also positive.
\end{proof}

\approxsumwu*
\begin{proof}
  Suppose that for some $\delta>0$, there is a $(1+\eps)$-approximation algorithm that solves the optimization version of \tf{\one}{}{\sumwu} in time $\Oh{(n+\frac{1}{2n\eps})^{2-\delta}}$. Since $nw_{\max}\geq\OPT$ for any given instance, setting $\eps:=\frac{1}{1+nw_{\max}}$ yields a solution with value $z$ such that
  \[\OPT\leq z\leq (1+\eps)\OPT=\OPT+\frac{\OPT}{1+nw_{\max}}<\OPT+1.\]
  Since all weights are integer, $\OPT$ is also integer and hence, $z=\OPT$. So we just solved the optimization version of \tf{\one}{}{\sumwu} exactly in time $\Oh{(n+\frac{1}{2n\eps})^{2-\delta}}=\Oh{(n+\frac{1+nw_{\max}}{2n})^{2-\delta}}\leq\Oh{(n+\frac{2nw_{\max}}{2n})^{2-\delta}}=\Oh{(n+w_{\max})^{2-\delta}}$. With that, we can also solve the decision problem in the same running time for any given threshold $y$ and by \cref{thm:exact:wjUj}, this refutes the $(\min,+)$-conjecture.
\end{proof}

\totaltardiness*
\begin{proof}
The NP-hardness of \tf{\one}{}{\sumt} is shown by Du and Leung~\cite{DL90}, who reduce from the NP-hard problem \textsc{Even-Odd-Partition} (or \textsc{EO-Partition} for short) via a restricted version thereof (\textsc{REO-Partition}).
While the hardness of \textsc{EO-Partition} is usually attributed to Garey and Johnson~\cite{GJ79}, the first reduction in the literature (to the best of our knowledge) is due to Garey, Tarjan and Wilfong~\cite{GTW88}.
We revisit the reductions in~\cite{GTW88} and~\cite{DL90} to prove~\cref{thm:1||sumT_j}.
\begin{problem}{\textsc{EO-Partition}}
\begin{myproblem}
\instance Integers $b_{1},\ldots,b_{2n}\in\mathbb{N}$ with $b_i > b_{i+1}$ for each $i\in [2n-1]$.
\task Decide whether there is a subset $S\subseteq [2n]$ such that $\sum_{i\in S}b_i = \sum_{i\in\bar{S}} b_i$ and $|S \cap \set{b_{2i-1},b_{2i}}| = 1$ for each $i\in[n]$.
\end{myproblem}
\end{problem}
In other words, the items $b_i$ are strictly decreasing and consist of $n$ pairs of items $(b_{2i-1},b_{2i})$, where the items of a pair may not be in the same partition.

The reduction from \PART to \textsc{EO-Partition} by to Garey, Tarjan and Wilfong~\cite{GTW88} is as follows: Given a \PART instance $a_{1},\ldots,a_{n}$, we may assume that $A=\sum_{i\in[n]} a_i$ is even because otherwise the instance is trivial.
We set $b_{2n} = 1$, $b_{2i - 1} = b_{2i} + a_i$ for each $i\in[n]$, and $b_{2i} = b_{2i + 1} + 1$ for each $i\in[n - 1]$. In other words, we start at the smallest item (which also has the largest index) and set it to $1$. Then we recursively define the other items, step by step: If we stay in the same pair $i$, we add $a_i$ and if we go from one pair to the next, we add only $1$. Hence, the items increase throughout the construction and we get $b_i > b_{i+1}$ for each $i\in [2n-1]$. Moreover, the difference between the larger item of a pair and the smaller one is $b_{2i - 1} - b_{2i} = a_i$ for every pair $i\in[n]$.

Suppose that the \PART instance is positive, i.e. we have a set $S\subseteq[n]$ such that $\sum_{i\in S}a_i=\sum_{i\in\overbar{S}}a_i$. Consider the set $T=\{j\in[2n]\,|\,(i\in S \land j=2i-1) \lor (i\in\overbar{S} \land j=2i)\}$, where we take all the odd-indexed (i.e. larger) items corresponding to items in $S$ and the even-indexed (i.e. smaller) items corresponding to items in $\overbar{S}$. It is not quite clear how large $\sum_{j\in T}b_j$ is, but using the fact that $b_{2i - 1} - b_{2i} = a_i$, we can see that
\begin{align*}
    \sum_{j\in T}b_j - \sum_{j\in \overbar{T}}b_j &= (\sum_{i\in S}b_{2i-1} + \sum_{i\in \overbar{S}}b_{2i}) - (\sum_{i\in S}b_{2i} + \sum_{i\in \overbar{S}}b_{2i-1}) \\
    &=\sum_{i\in S}(b_{2i-1}-b_{2i}) + \sum_{i\in \overbar{S}}(b_{2i}-b_{2i-1}) \\
    &=\sum_{i\in S}a_i - \sum_{i\in \overbar{S}}a_i \\
    &=0
\end{align*}
and hence, $T$ and $\overbar{T}$ are a valid partition.

For the other direction, suppose $T$, $\overbar{T}$ is a solution of the \textsc{EO-Partition} instance. Define a solution of the corresponding \PART instance as follows: $S=\{i\in[n]\,|\,b_{2i-1}\in T\}$. Using essentially the same transformations as above, it follows that $\sum_{i\in S}a_i - \sum_{i\in \overbar{S}}a_i=\sum_{j\in T}b_j - \sum_{j\in \overbar{T}}b_j$, which is equal to zero, by assumption. So $S$ is indeed a solution of the \PART instance.

Note that $\sum_{j\in[2n]} b_j = \Oh{n^2 + nA}$, since the largest item $b_1$ is bounded by $\Oh{n+A}$.
Furthermore, note that in the resulting \textsc{EO-Partition} instance we have $\sum_{i\in[n]}(b_{2i-1} - b_{2i}) = \sum_{i\in[n]} a_i$ and therefore may assume that this number is even in the following.

The \textsc{REO-Partition} problem was introduced by Du and Leung~\cite{DL90} and is very similar to \textsc{EO-Partition}.
However, in this version of the problem the input consist of integers $c_{1},\ldots,c_{2n}\in\mathbb{N}$ with $c_i > c_{i+1}$ for each $i\in [2n-1]$, $c_{2i} > c_{2i + 1} + \delta$ for each $i\in[n-1]$, and $c_i > n(4n + 1)\delta + 5n(c_1 - c_{2n})$ for each $i\in[2n]$, where $\delta = \frac{1}{2}\sum_{i\in[n]}(c_{2i-1} - c_{2i}) $. So in this restricted variant, the items of subsequent pairs have difference $>\delta$ and each item is larger than some value depending on $n$, $\delta$ and the difference between the largest and smallest item.
The reduction by Du and Leung~\cite{DL90} from \textsc{EO-Partition} to \textsc{REO-Partition} is as follows: Let $\Delta = \frac{1}{2}\sum_{i\in[n]}(b_{2i-1} - b_{2i})=\frac{1}{2}A$ and
\begin{alignat*}{3}
    &c_{2i - 1} &&= b_{2i - 1} &+ (9n^2 + 3n - (i-1))\Delta + 5n(b_1 - b_{2n}) \\
    &c_{2i} &&= b_{2i} &+ (9n^2 + 3n - (i-1))\Delta + 5n(b_1 - b_{2n})
\end{alignat*}
for every $i\in[n]$. We know $\Delta\in\mathbb{N}$ as the sum is even, by a previous assumption. Note that $\delta = \Delta$ since $c_{2i-1} - c_{2i} = b_{2i-1} - b_{2i}$ holds for each $i\in[n]$.
Moreover, $c_i > c_{i+1}$ for each $i\in [2n-1]$ since $b_i > b_{i+1}$. We also have $c_{2i} > c_{2i + 1} + \delta$ for each $i\in[n-1]$, since we get an additional $\Delta=\delta$ between subsequent pairs and since already $b_{2i} > b_{2i + 1}$. Finally, we also get $c_i > n(4n + 1)\delta + 5n(c_1 - c_{2n})$ for each $i\in[2n]$, as we will show. First note that
\begin{align*}
    c_1-c_{2n}&=b_{1} + (9n^2 + 3n - (1-1))\Delta + 5n(b_1 - b_{2n}) \\
    &\,\,\,- (b_{2n} + (9n^2 + 3n - (n-1))\Delta + 5n(b_1 - b_{2n})) \\
    &=b_1-b_{2n}+(n-1)\Delta
\end{align*}
and hence:
\begin{align*}
    n(4n + 1)\delta + 5n(c_1 - c_{2n}) &=n(4n + 1)\Delta + 5n(b_1-b_{2n}+(n-1)\Delta) \\
    &=n(4n + 1)\Delta + 5n(n-1)\Delta + 5n(b_1-b_{2n}) \\
    &=(n(4n + 1) + 5n(n-1))\Delta + 5n(b_1-b_{2n}) \\
    &=(4n^2+n+5n^2-5n)\Delta + 5n(b_1-b_{2n}) \\
    &=(9n^2-4n)\Delta + 5n(b_1-b_{2n}) \\
    &<b_{2n}+(9n^2+2n+1)\Delta + 5n(b_1-b_{2n}) \\
    &=b_{2n}+(9n^2+3n-(n-1))\Delta + 5n(b_1-b_{2n}) \\
    &=c_{2n}
\end{align*}
So the inequality holds for the smallest item, which means that it holds for all items. We can conclude that the constructed instance of \textsc{REO-Partition} is valid. 

It is not hard to verify that a solution of the \textsc{EO-Partition} instance can be transformed to a solution of the \textsc{REO-Partition} instance and vice-versa by just selecting the corresponding $c_j$ (respectively $b_j$).

If we consider the two reductions one after another and use previously observed bounds for $b_1$, $\sum_{j\in[2n]} b_j$ and $\Delta$, we get: 
\begin{align*}
\sum_{j\in[2n]} c_i &=\sum_{j\in[2n]} (b_j+(9n^2 + 3n-(j-1))\Delta + 5n(b_1 - b_{2n})) \\
&\leq 2n((9n^2 + 3n)\Delta + 5n(b_1 - b_{2n})) + \sum_{j\in[2n]} b_j\\
&= \Oh{n^3\Delta} + \Oh{n^2b_1} + \sum_{j\in[2n]} b_j\\
&= \Oh{n^3A} + \Oh{n^3+n^2A} + \Oh{n^2 + nA}\\
&= \Oh{n^3A}
\end{align*}

In the last step, Du and Leung construct a \tf{\one}{}{\sumt} instance from the \textsc{REO-Partition} instance. For the details, see the paper~\cite{DL90}. What is important for our lower bound are the parameter sizes of the \tf{\one}{}{\sumt} instance, namely the number of jobs and the total processing time. The due dates are trivially bounded by the total processing time, which makes them less interesting in a lower bound. An examination of the construction (page 487) gives the following parameters:
\begin{itemize}
    \item $N:=3n+1$ jobs and
    \item total processing time $K:=P=\sum_{i\in[2n]} c_i + (n+1)(4 n + 1)\delta = \Oh{n^3A}=\Oh{\text{poly}(n)A}$.
\end{itemize}
Since the reduction is polynomial in $n$, we can use \cref{lem:reductions} to conclude the proof.
\end{proof}

\reject*
\begin{proof}
  Consider an instance $a_1,\dots,a_n$, $T$ of \SSS. We create an instance of \tf{\one}{\rej}{\cmax} $n$ jobs, $p_j=w_j=a_j$ for every $j\in[n]$, $y=T$ and $Q=A-T$. We can now see that there is a subset of items that sums up to $T$, if and only if there is a subset of jobs that is scheduled in time $T$, the rest of the jobs is rejected and total weight of rejected jobs is at most $A-T$.
  
  In the constructed instance, we have $N:=n$ jobs and using $K:=y+P+Q+W=T+A+A-T+A=\Oh{A}$ as parameter with \cref{lem:reductions} proves the claim, since the reduction takes time $\Oh{n}$.
\end{proof}

\twocmax*
\begin{proof}
We show that the lower bound $\Oh{2^{\delta n}A^{1-\eps}}$ for \PART can be transferred to \tf{\two}{}{\cmax}.
Let $a_1, \dots, a_n$ be a \PART instance. Construct the \tf{\two}{}{\cmax} instance by setting $p_j = a_j$ for every $j \in [n]$ and $y = \frac{1}{2}A$. It is easy to see that there is a partition of the items into two subsets of equal sum, if and only if the jobs can be split among the two machines such that each one gets assigned jobs with total processing time $y$.

Since the constructed instance has $N:=n$ jobs and takes time linear in $n$, we can prove the claim by using $K:=y+P=\frac{1}{2}A+A=\Oh{A}$ as parameter in \cref{lem:reductions}.
\end{proof}

\fixedcmax*
\begin{proof}
We assume that $m>1$, as the problem is trivial on a single machine. In~\cite{CJZ14}, Chen \etal show that the known approximation schemes for \tf{\fixed}{}{\cmax} are essentially optimal. In particular, they also show that there is no $2^{\Oh{m^{\frac{1}{2}-\delta}\sqrt{|I|}}}$-time exact algorithm for any $\delta>0$, unless the ETH fails. This is done by a reduction from \textsc{3-Sat} via \textsc{3-Dimensional-Matching} to \tf{\fixed}{}{\cmax}. 

For us, the crucial part about these reductions is that we can choose $m$ arbitrarily and if the original \textsc{3-Sat} formula has $n'$ variables, the \tf{\fixed}{}{\cmax} instance has $\Oh{n'+m}$ jobs and total processing time bounded by $P\leq n'm^{\Oh{\frac{n'\log(m)}{m}}}$.

This can be seen in the paper by Chen \etal~\cite{CJZ14} on page 666, where a job is constructed for each of the $\Oh{n'}$ matches and elements in addition to at most $m$ dummy jobs and one huge job. The processing time of the huge job is set to $6m^3(m+1)\sum_{i=1}^{9q+\tau}\alpha^i$ minus the total processing time of the other constructed jobs, where $q=\frac{n'}{m}$, $\tau=\Oh{\frac{n'\log(m)}{m}}$ and $\alpha=6m^4+6m^3+6m^2$. Note that in the reduction it is assumed that $q$ is integer, so $n'\geq m$. This is achieved by adding dummy elements to the \textsc{3-Dimensional-Matching} instance. Hence, the total processing time is equal to $6m^3(m+1)\sum_{i=1}^{9q+\tau}\alpha^i$, which can be bounded by
\[n'm^{\Oh{\frac{n'\log(m)}{m}}}=2^{\log(n')}2^{\Oh{\frac{n'\log^2(m)}{m}}}=2^{\Oh{\log(n')+\frac{n'\log^2(m)}{m}}}=2^{\Oh{\frac{n'\log^2(m)}{m}}},\]
using
\begin{align*}
    m\leq n' \implies \frac{m}{\log(m)}\leq \frac{n'}{\log(n')} &\implies \frac{m}{\log^2(m)}\leq \frac{n'}{\log(n')}\\
    &\iff \log(n')\leq\frac{n'\log^2(m)}{m},
\end{align*}
where the first implication follows because the function $\frac{k}{\log(k)}$ is monotone for values $k\geq2$ (and we assumed $m$ -- and hence also $n'$ -- to be larger than $1$).

Now, suppose that we have an $\Oh{nmP^{\oh{\frac{m}{\log^2(m)}}}}$-time algorithm solving \tf{\fixed}{}{\cmax}. Using the reduction by Chen \etal~\cite{CJZ14}, which has a running time $\text{poly}(n)=n^c$, we could then solve \textsc{3-Sat} in time:
\begin{align*}
    \Oh{n^c+nmP^{\oh{\frac{m}{\log^2(m)}}}} &\leq 2^{\Oh{\log(n')}}\left(2^{\Oh{\frac{n'\log^2(m)}{m}}}\right)^{\oh{\frac{m}{\log^2(m)}}}\\
    &\leq 2^{\Oh{\log(n')}}2^{\oh{n'}}\\
    &=2^{\Oh{\log(n')}+\oh{n'}}\\
    &=2^{\oh{n'}}
\end{align*}
This contradicts the ETH and proves the theorem.
\end{proof}

\ptwosumwc*
\begin{proof}
    We only show the correctness of the reduction, here. The fact that the lower bound follows from the reduction has already been shown. Remember that given a \PART instance $a_1, \dots, a_n$, we construct a \tf{\two}{}{\sumwc} instance in the following way: Let $p_j = w_j = a_j$ for each $j\in[n]$ and set the limit $y = \sum_{1\leq i \leq j \leq n} a_j a_i - \frac{1}{4}A^2$. 
    
	Note that the execution order of the jobs on one specific machine does not influence the sum of weighted completion times for that machine, since $p_j=w_j$ for all jobs $j$: This follows from the observation that $\sum_{j\in[k]} w_j C_j = \sum_{j\in[k]} p_j (\sum_{i\leq j} p_i) = \sum_{1 \leq i \leq j \leq k} p_j p_i$ holds for any $k$ jobs running on one machine.
	Before we prove the correctness of the construction, consider a schedule of jobs $\{1,\hdots,n\}=S\cup\overbar{S}$ with $w_j=p_j$, where jobs in $S$ are scheduled on the first machine and jobs in $\overbar{S}$ are scheduled on the second machine. We wish to show that the total weighted completion time of this schedule is $\sum_{j=1}^nw_jC_j-\sum_{j\in S}w_j\sum_{j\in \overbar{S}}p_j=\sum_{1\leq i\leq j\leq n} p_j p_i-\sum_{j\in S}w_j\sum_{j\in \overbar{S}}p_j$. 
	If all jobs were scheduled on the first machine, the value would be just $\sum_{1\leq i\leq j\leq n} p_j p_i$, as argued above. Again, this does not depend on the order of the jobs, but let us assume that the jobs in $\overbar{S}$ are scheduled first. Moving them to the second machine reduces the total weighted completion time by $\sum_{j\in S}w_j\sum_{j\in \overbar{S}}p_j$, as the completion time of each job in $S$ is reduced by $\sum_{j\in \overbar{S}}p_j$ and the completion time of the jobs in $\overbar{S}$ stays the same. 	
	
	Now, consider a solution $S\cup\overbar{S}$ of \PART. We schedule the jobs corresponding to items in $S$ on the first machine and the rest on the second machine, in arbitrary order. By the previous observations, this schedule has total weighted completion time:
	\begin{align*}
	    \sum_{1\leq i\leq j\leq n} p_j p_i-\sum_{j\in S}w_j\sum_{j\in \overbar{S}}p_j &=\sum_{1\leq i\leq j\leq n} a_j a_i-\sum_{j\in S}a_j\sum_{j\in \overbar{S}}a_j\\
	    &=\sum_{1\leq i\leq j\leq n} a_j a_i-\frac{1}{2}A\frac{1}{2}A \\
	    &=\sum_{1\leq i\leq j\leq n} a_j a_i-\frac{1}{4}A^2
	\end{align*}
	So this schedule meets the threshold $y=\sum_{1\leq i\leq j\leq n} a_j a_i-\frac{1}{4}A^2$. 
	
	For the other direction, it is only important to see that we have the term $\sum_{1\leq i\leq j\leq n} a_j a_i$ in the total weighted completion time no matter what and that the selection of jobs $\overbar{S}$ we put on the second machine determines the second part $\sum_{j\in S}a_j\sum_{j\in \overbar{S}}a_j$ that is subtracted from the first term. Since the total sum $A=\sum_{i=1}^na_i=\sum_{j\in S}a_j+\sum_{j\in \overbar{S}}a_j$ is fixed, the maximum of the product $\sum_{j\in S}a_j\sum_{j\in \overbar{S}}a_j$ is attained when $\sum_{j\in S}a_j=\sum_{j\in \overbar{S}}a_j$. This is also exactly the case where the value is equal to $\frac{1}{4}A^2$. In all other cases, the product is smaller and hence subtracting less from $\sum_{1\leq i\leq j\leq n} a_j a_i$ gives us an objective value larger than $y$. With this observation, clearly the jobs must be split such that the jobs on machine 1 have the same total processing time as the jobs on machine 2, which is only possible if the corresponding \PART instance is positive.
\end{proof}

\twosize*
\begin{proof}
	We show that the lower bound $\Oh{2^{\delta n}(y+P)^{1-\eps}}$ for \tf{\two}{}{\cmax} can be transferred to \tf{\two}{\size}{\cmax}.
	Construct the \tf{\two}{\size}{\cmax} instance from the \tf{\two}{}{\cmax} instance by setting the size to $1$ for each job. The correctness for this reduction is trivial and neither the number of jobs nor the total processing time or threshold changes, so the lower bound for \tf{\two}{}{\cmax} directly applies to \tf{\two}{\size}{\cmax} and we can conclude that there is no algorithm that solves \tf{\two}{\size}{\cmax} in $\Oh{2^{\delta n}(y+P)^{1-\eps}}$, unless the SETH fails.
\end{proof}

\twoany*
\begin{proof}
	We show that the lower bound $\Oh{2^{\delta n}(y+P)^{1-\eps}}$ for \tf{\two}{}{\cmax} can be transferred to \tf{\two}{\any}{\cmax}.	
	Construct the \tf{\two}{\any}{\cmax} instance by setting $p_{j,1} = p_{j,2}=p_j$ for every $j\in[n]$. Any schedule for \tf{\two}{}{\cmax} also represents a schedule for \tf{\two}{\any}{\cmax} and since no advantage can be achieved by scheduling any job on two machines, a feasible schedule for \tf{\two}{\any}{\cmax} is also feasible for \tf{\two}{}{\cmax}. 
	Again, this does not change the size of the instance and hence the lower bound for \tf{\two}{}{\cmax} directly applies to \tf{\two}{\any}{\cmax} so we can conclude that there is no algorithm that solves the problem in time $\Oh{2^{\delta n}(y+P)^{1-\eps}}$, unless the SETH fails.
\end{proof}

\dynprog*
\begin{proof}
For three machines, the main idea of the dynamic program stays the same; the recurrence formula just becomes a bit more complicated. We create a field $F(j,t_1,t_2)$, which tells us the minimum load we can get on machine 3, if we schedule all the jobs in $[j]$ such that machine 1 and 2 have load $t_1$ and $t_2$, respectively. In analogy to the dynamic program for two machines, we define the recurrence formula:
\begin{align*}
    F(j,t_1,t_2)= \min
    \begin{cases}
        T(j-1,t_1-p_j(1),t_2) \\
        T(j-1,t_1,t_2-p_j(1)) \\
        T(j-1,t_1,t_2)+p_j(1) \\
        T(j-1,t_1-p_j(2),t_2-p_j(2)) \\
        T(j-1,t_1-p_j(2),t_2)+p_j(2) \\
        T(j-1,t_1,t_2-p_j(2))+p_j(2) \\
        T(j-1,t_1-p_j(3),t_2-p_j(3))+p_j(3)
    \end{cases}
\end{align*}
The cases correspond to scheduling job $j$ on machine 1, on machine 2, on machine 3, on machine 1 and 2, on machine 1 and 3, on machine 2 and 3 and finally on all three machines. Again, the initial entries are $F(0,0,0)=0$ and $F(0,t_1,t_2)=\infty$ for every $t_1,t_2\in[y]$.

This time, we need to compute $\Oh{nP^2}$ entries; the actual distribution of our jobs among the machines can be obtained in the standard way, i.e. by remembering how we obtained every entry or by going backwards through the table. However, in this case, it is not directly clear that this distribution of jobs yields a feasible schedule with makespan at most $y$. Fortunately, as Du and Leung~\cite{DL89} observed, there is a canonical schedule in the case of three machines: The jobs are swapped such that there are only two-machine jobs on machines 1 and 2 and on machines 2 and 3. Then, the three-machine jobs are moved to the beginning of the schedule, followed by the two-machine jobs on machine 1 and 2. Finally, the two-machine jobs on machine 2 and 3 are executed at the end of the schedule and the one-machine jobs are executed in between. Using this canonical schedule and the distribution of the jobs to machines, we can obtain the actual schedule (i.e. with starting times). 
\end{proof}

\section{Strongly NP-Hard Problems}\label{app:strongly}
In this section, we show all our SETH-based lower bounds for strongly NP-hard problems. It is important to note that unless P=NP, these problems cannot have pseudo-polynomial algorithms~\cite{CPW98}. However, our lower bounds do not only exclude pseudo-polynomial algorithms; algorithms with a super-polynomial but sub-exponential dependency on $n$ (and a linear dependency on the other parameters) are also impossible under SETH. So even though these results are not as strong as those for weakly NP-hard problems, they might still be of interest for parameterized or approximation algorithms.

\subsection*{Jobs With Due Dates}
Our lower bound from \cref{thm:1||sumT_j} for \tf{\one}{}{\sumt} also implies a lower bound for \tf{\one}{}{\sumwt} (see \cref{cor:1||sumw_jT_j}). But with a more elaborate reduction that actually uses (though only two) different weights, we get a stronger lower bound for \tf{\one}{}{\sumwt}, the problem of minimizing the total weighted tardiness on a single machine.
\begin{theorem}\label{thm:1||sumw_jT_j}
	For every $\eps>0$, there is a $\delta>0$ such that \tf{\one}{}{\sumwt} cannot be solved in time $\Oh{2^{\delta n} \left(d_{\max}+P+W+\sqrt{y}\right)^{1 - \eps}}$, unless the SETH fails.
\end{theorem}
\begin{proof}
	We revisit the reduction by Lenstra \etal\cite{LRB77} from \SSS to \tf{\one}{}{\sumwt}.
	
	Consider a \SSS instance $a_1, \dots, a_n$, $T$. We set $N = n + 1, p_j = w_j = a_j, d_j = 0$ for each $j \in [n]$, $p_N = 1, w_N = 2, d_N = T + 1$, and $y = \sum_{1 \leq i \leq j \leq n} a_i a_j + A - T$.  Again, the idea is that the newly added job $n$ acts as a barrier at time $T$ and the solution to the \SSS instance mapped to the \tf{\one}{}{\sumwt} instance and job $N$ have to be scheduled before $T + 1$.
	
	Suppose the \SSS instance is positive, i.e. there is a subset $S$ of items summing up to exactly $T$. Schedule the jobs corresponding to $S$ first, then job $N$ and finally the rest of the jobs. If we ignore job $N$ for a moment, we have only the $n$ jobs with $p_j=w_j=a_j$, so their total weighted completion time is equal to $\sum_{1 \leq i \leq j \leq n} a_i a_j$, which is also the total weighted tardiness, since the due dates of these jobs are all zero. Now, adding job $N$ to the schedule at time $T$ increases the tardiness of every job in $\overbar{S}$ by $1$. This increase by $1$ is multiplied by the weight of each job in $\overbar{S}$ and we get a total increase of $A-T$, since that is the sum of the weights of the jobs in $\overbar{S}$. So the total weighted tardiness of the constructed schedule is equal to $\sum_{1 \leq i \leq j \leq n} a_i a_j + A-T=y$ and hence the instance is positive.
	
	For the other direction, assume that we are given a schedule with total weighted tardiness at most $y$. Consider two cases: 
	If job $N$ is scheduled after $T$, say at time $T+k$ with $k>0$, then the minimum total weighted tardiness is achieved by having no gaps in the schedule, i.e. there are jobs with total processing time $T+k$ scheduled before job $N$ and jobs with total processing time $A-T-k$ are scheduled after. Now, the total weighted tardiness of this schedule is $\sum_{1 \leq i \leq j \leq n} a_i a_j + (A-T-k)+2k=\sum_{1 \leq i \leq j \leq n} a_i a_j + A-T+k>y$, since the jobs after job $N$ are delayed by one time unit and have total weight $A-T-k$ and job $N$ is late by $k$ time units and has weight $2$. This is a contradiction, so job $N$ has to be scheduled at time $T$.
	Now we know from the observations in the proof of the first direction that a gap-less schedule with job $N$ scheduled at time $T$ has total weighted tardiness exactly $y$. And if there is a gap, the total weighted tardiness strictly increases. Hence, we can conclude that there can be no gap and that the jobs scheduled before job $N$ have a total processing time exactly $T$ and the corresponding items form a solution of the \SSS instance.
	
	Since the constructed instance has $N=n+1$ jobs and moreover $d_{\max}=T+1$, $P=A+1$, $W=A+2$ and 
	\begin{align*}
	    y=\sum_{1 \leq i \leq j \leq n} a_i a_j + A - T\leq\sum_{i=1}^n\left(a_i\sum_{j=1}^n a_j\right) + \Oh{T} &= \sum_{i=1}^n\left(a_iA\right) + \Oh{T} \\
	    &\leq \Oh{A^2} + \Oh{T} \\
	    &\leq \Oh{T^2}
	\end{align*}
	setting $K:=d_{\max}+P+W+\sqrt{y}=T+1+A+1+A+2+\Oh{T}=\Oh{A}$ and using \cref{lem:reductions} finishes the proof. Note that the reduction is polynomial in $n$.
\end{proof}

\subsection*{Jobs With Release Dates}

We consider the problem \tf{\one}{\rj}{\sumwc}, where we aim to minimize the total weighted completion time subject to having release dates $r_j$ for every job. With a classical reduction from \SSS, we get the following lower bound:

\begin{theorem}\label{thm:1|r_j|sumw_jC_j}
	For every $\eps>0$, there is a $\delta>0$ such that \tf{\one}{\rj}{\sumwc} cannot be solved in time $\Oh{2^{\delta n} \left(r_{\max}+P+W+\sqrt{y}\right)^{1 - \eps}}$, unless the SETH fails.
\end{theorem}

\begin{proof}
    We revisit the reduction by Rinnooy Kan~\cite{Kan12}, who reduces from \SSS. Let $a_1, \dots, a_n$, $T$ be a \SSS instance.
	Construct a \tf{\one}{\rj}{\sumwc} instance by setting $N = n + 1$, $p_j = w_j = a_j, r_j = 0$ for each $j \in [N-1]$ and $p_N = 1, w_N~=~2, r_N~=~T,$ $y~=~\sum_{1 \leq i \leq j \leq n} a_i a_j + A + T + 2$.
	The idea is that the split job $N$ has to be scheduled at its release date $T$ and there cannot be any gaps in the schedule. The job $N$ then acts as a barrier between jobs corresponding to items from the \SSS solution and the rest. 
	
	For the first direction, assume that there is a subset $S$ that is a solution of our \SSS instance. Schedule the jobs corresponding to items in $S$ before $T$, then schedule the split job $N$ and finally the rest of the jobs. If we ignore the split job for a second, the total weighted completion time of this schedule is $\sum_{1 \leq i \leq j \leq n} a_i a_j$, regardless of the order of jobs (this follows again from the fact that $p_j=w_j$ for all the jobs). Now, if we add the split job, we get its completion time $T+1$, multiplied with its weight $2$; moreover, all jobs scheduled after it are delayed by $1$. These delays are in turn multiplied by the weights (which are equal to their processing times). Hence, the total weighted completion time becomes:
	\[\sum_{i=1}^Nw_jC_j=\sum_{1 \leq i \leq j \leq n} a_i a_j + 2(T+1)+A-T=\sum_{1 \leq i \leq j \leq n} a_i a_j+A+T+2=y\]
	So our constructed schedule meets the target $y$ and is therefore feasible.
	
	For the other direction, consider any schedule for the constructed instance with total weighted completion time at most $y$ and distinguish two cases: 
	
	If the split job $N$ is scheduled directly at time $T$, there cannot be a gap before job $N$. Otherwise, the weighted completion time is
	\[\sum_{i=1}^Nw_jC_j\geq\sum_{1 \leq i \leq j \leq n} a_i a_j+2(T+1)+A-T+1>y,\]
	since the load of the jobs scheduled after job $N$ is at least $A-T+1$ because of the gap. So this sub-case leads to a contradiction. If there is no gap, the jobs scheduled before job $N$ have total processing time exactly $T$ and the corresponding items are a solution of the original \SSS instance.
	
	For the second case, assume that job $N$ starts after its release date $T$, say at time $T+k$, where $k>0$. Without loss of generality, we can also assume that there is no gap in the schedule before the execution of job $N$, since such a gap would only increase the weighted completion time. So there are jobs with processing time $T+k$ scheduled before $N$ and jobs with processing time $A-(T+k)$ scheduled after job $N$. Thus, the total weighted completion time of the schedule is
	\[\sum_{i=1}^Nw_jC_j\geq\sum_{1 \leq i \leq j \leq n} a_i a_j+2(T+k+1)+A-(T+k)=\sum_{1 \leq i \leq j \leq n} a_i a_j+A+T+2+k,\]
	which is strictly larger than $y$, a contradiction. 
	
	As job $N$ cannot be scheduled before $T$ because of its release date, we have to end up in the first case, where we find a solution of the original \SSS instance.
	
	By construction, we have $r_{\max} = T$, $P=A+1$ and $W=A+2$. Moreover, we show that $y\leq\Oh{T^2}$:
	\begin{align*}
	    y=\sum_{1 \leq i \leq j \leq n} a_i a_j + A + T + 2 \leq \sum_{i=1}^n\left(a_i\sum_{j=1}^n a_j\right)+\Oh{T} &= \sum_{i=1}^n a_iA+\Oh{T}\\
	    &\leq\Oh{A^2+T}\\
	    &=\Oh{T^2}
	\end{align*}
	Now, since the constructed instance has $N=n+1$ jobs and the reduction is polynomial in $n$, setting $K:=r_{\max}+P+W+\sqrt{y}=T + A + 1 + A + 2 + \Oh{T}=\Oh{A}$ and using \cref{lem:reductions} proves the claim.
\end{proof}

A similar idea also works for \tf{\one}{\rj}{\tmax}, where we aim to minimize the maximum tardiness and have additional release dates. Again, a classical reduction from \SSS gives us a lower bound:

\begin{theorem}\label{thm:1|r_j|T_{max}}
	For every $\eps>0$, there is a $\delta>0$ such that \tf{\one}{\rj}{\tmax} cannot be solved in time $\Oh{2^{\delta n} (d_{\max} + r_{\max} + P + y)^{1 - \eps}}$, unless the SETH fails.
\end{theorem}

\begin{proof}
	Lenstra \etal~\cite{LRB77} show the NP-hardness of \tf{\one}{\rj}{\lmax} by a reduction from \KS (which is a generalization of \SSS: If for each item weight and profit are the same, the problems are equivalent). We revisit this reduction to prove~\cref{thm:1|r_j|T_{max}}. Note that the reduction by Lenstra \etal is supposedly for the $L_{\max}$-version, but the same reduction also works for $T_{\max}$. This is because $T_{\max}$ and $L_{\max}$ are the same in the constructed instance, since there is a job with $d_j=r_j+p_j$. Hence, the maximum lateness cannot be negative and has to be equal to the maximum tardiness.
	
	Consider a \SSS instance $a_1, \dots, a_n$, $T$.
	We set $N = n + 1, r_j = 0,$ $p_j~=~a_j,$ $d_j~=~A + 1$ for each $j \in [n]$, $r_N = T, p_N = 1, d_N = T + 1$, and $y = 0$.
	Once more, the idea is that the newly added job $N$ acts as a barrier at time $T$ and the solution to the \SSS instance mapped to the \tf{\one}{\rj}{\tmax} instance has to be scheduled before $T$.
	
	For the first direction, assume we have a subset $S$ of items summing up to $T$. Then we can schedule all the jobs corresponding to the items in $S$ before $T$ (with processing time $T$), then the job $N$ (with processing time 1) and then the rest of the jobs (with total processing time $A-T$). So we get a schedule in which each job makes its due- and release date, i.e. one with objective value $y=0$.
	
	For the other direction, assume we are given a schedule with objective value $y\leq0$. This means that no job can be late, which has two consequences: Job $N$ with processing time 1 has to be scheduled exactly at time $T$ to meet its release date $T$ and its due date $T+1$. Moreover, all other jobs have to be finished before their uniform due date $A+1$. Since their total processing time is equal to $A$ and the remaining space is also equal to $A$, there can be no gap in the schedule. Instead, the jobs are perfectly divided into a set of jobs $S$ that are scheduled before time $T$ and the rest of the jobs. So the jobs in $S$ have total processing time $T$ and and correspond to a subset of the items summing up to $T$ in the original \SSS instance.
	
	Since $N=n+1$, $r_{\max} = T$, $y=0$ and $P=d_{\max}=A+1$, using parameter $K:=d_{\max} + r_{\max} + P + y=A+1 + T + A + 1 + 0=\Oh{A}$ with \cref{lem:reductions} proves the claim. Note that the reduction is linear in $n$.
\end{proof}

\subsection*{Jobs With Deadlines}

In the problem \tf{\one}{\dj}{\sumwc}, we aim to minimize the total weighted completion time subject to deadlines $d_j$. With a classical reduction from \SSS, we get the following result:

\begin{theorem}\label{thm:1|d_j|sumw_jC_j}
	For every $\eps>0$, there is a $\delta>0$ such that \tf{\one}{\dj}{\sumwc} cannot be solved in time $\Oh{2^{\delta n} \left(d_{\max}+P+W+\sqrt{y}\right)^{1 - \eps}}$, unless the SETH fails.
\end{theorem}

\begin{proof}
    We revisit the reduction by Lenstra \etal~\cite{LRB77} from \SSS to \tf{\one}{\dj}{\sumwc}.
    
	Consider a \SSS instance $a_1, \dots, a_n$, $T$.
	We set $N = n + 1, p_j = w_j = a_j, d_j = A + 1$ for each $j \in [n]$, $p_N = 1,$ $w_N~=~0,$ $d_N~=~T+1$, and $y = \sum_{1 \leq i \leq j \leq n} a_i a_j + A - T$.
	Once again, the idea is that the newly added job $N$ acts as a barrier at time $T$ and the solution to the \SSS instance mapped to the \tf{\one}{\dj}{\sumwc} instance and job $N$ have to be scheduled before $T + 1$.
	
	Suppose we are given a subset $S$ of items summing up to exactly $T$. Then we can schedule the corresponding jobs first, then job $N$ and finally the rest of the jobs. All jobs meet their deadline, the total weighted completion time without job $N$ is $\sum_{1 \leq i \leq j \leq n} a_i a_j$ and adding job $N$ delays the later jobs by one time unit, resulting in an increase of $A-T$ in the total weighted completion time, since that is that is the total weight of these jobs. Hence, we have a feasible schedule with total weighted completion time $\sum_{1 \leq i \leq j \leq n} a_i a_j+A-T=y$.
	
	For the other direction, suppose we are given a schedule with total weighted completion time at most $y$ and consider two cases: 
	If job $N$ is scheduled at time $T-k$ with $k>0$, even a gap-less schedule has total weighted completion time $\sum_{1 \leq i \leq j \leq n} a_i a_j+(A-T+k)>y$, since now jobs with total weight $A-T+k$ are delayed by the job $N$. 
	If -- on the other hand -- job $N$ is scheduled at time $T$, a gap-less schedule has total weighted completion time $y$ (see above) and \emph{only} a gap-less one. Hence, we get a subset of jobs with total completion time exactly $T$, which corresponds to a solution of the \SSS instance.
	
	Since the reduction is polynomial in $n$ and we have $N=n+1$, $d_{\max}=T+1$, $P=A+1$, $W=A$ and $\sqrt{y}=\Oh{T}$ (see the proof of~\cref{thm:1||sumw_jT_j}), using $K:=d_{\max}+P+W+\sqrt{y}=T+1+A+1+A+\Oh{T}=\Oh{A}$ as parameter with \cref{lem:reductions} proves the claim.
\end{proof}

\section{Implications for Other Objective Functions}\label{app:implications}
In this section, we make use of the fact that the common objective functions are partially ordered in complexity. Using classical reductions, we can transfer our lower bounds to a wide range of other scheduling problems. 

\begin{figure}
    \centering
    \scalebox{0.885}{
    \begin{tikzpicture}
        \tikzset{minimum width=1.5cm, minimum height=0.7cm}
    
        \node[draw] (Cmax) at (0,0) {$C_{\max}$};
        \node[draw] (Lmax) at (5,0) {$L_{\max}$};
        \node[draw] (Tmax) at (3,0) {$T_{\max}$};
        \node[draw] (SumC) at (5,4) {$\sum C_j$};
        \node[draw] (SumWC) at (9,4) {$\sum w_jC_j$};
        \node[draw] (SumU) at (9,0) {$\sum U_j$};
        \node[draw] (SumWU) at (12,0) {$\sum w_jU_j$};
        \node[draw] (SumT) at (9,2) {$\sum T_j$};
        \node[draw] (SumWT) at (12,2) {$\sum w_jT_j$};
        \node[draw] (SumV) at (9,-2) {$\sum V_j$};
        \node[draw] (SumWV) at (12,-2) {$\sum w_jV_j$};
        \node[draw] (Fmax) at (3,-2) {$F_{\max}$};
        \node[draw] (SumF) at (9,6) {$\sum F_j$};
        \node[draw] (SumWF) at (12,6) {$\sum w_jF_j$};
        
        \draw[->] (Cmax) -- node[midway,above,sloped] {\small $d_j:=0$} (Tmax);
        \draw[->] (Tmax) -- (Lmax);
        \draw[->] (Cmax) -- node[midway,above,sloped] {\small $r_j:=0$} (Fmax);
        \draw[->] (Lmax) -- node[midway,above,sloped] {\small $d_j:=d_j+y$} (SumU);
        \draw[->] (Lmax) -- node[midway,above,sloped] {\small $d_j:=d_j+y$} (SumV);
        \draw[->] (Lmax) -- node[midway,above,sloped] {\small $d_j:=d_j+y$} (SumT);
        \draw[->] (SumU) -- node[midway,above,sloped] {\small $w_j:=1$} (SumWU);
        \draw[->] (SumV) -- node[midway,above,sloped] {\small $w_j:=1$} (SumWV);
        \draw[->] (SumT) -- node[midway,above,sloped] {\small $w_j:=1$} (SumWT);
        \draw[->] (SumC) -- node[midway,above,sloped] {\small $w_j:=1$} (SumWC);
        \draw[->] (SumC) -- node[midway,above,sloped] {\small $r_j:=0$} (SumF);
        \draw[->] (SumF) -- node[midway,above,sloped] {\small $w_j:=1$} (SumWF);
        \draw[->] (SumWC) -- node[midway,above,sloped] {\small $d_j:=0$} (SumWT);
        \draw[->] (SumC) -- node[midway,above,sloped] {\small $d_j:=0$} (SumT);
        \draw[->] (SumWC) -- node[midway,above,sloped] {\small $r_j:=0$} (SumWF);
    \end{tikzpicture}
    }
    \caption{Classical reductions between objective functions (see e.g.~\cite{LLRS89} and the very useful website \url{http://schedulingzoo.lip6.fr/about.php}).}
    \label{fig:objectivefunctions}
\end{figure}
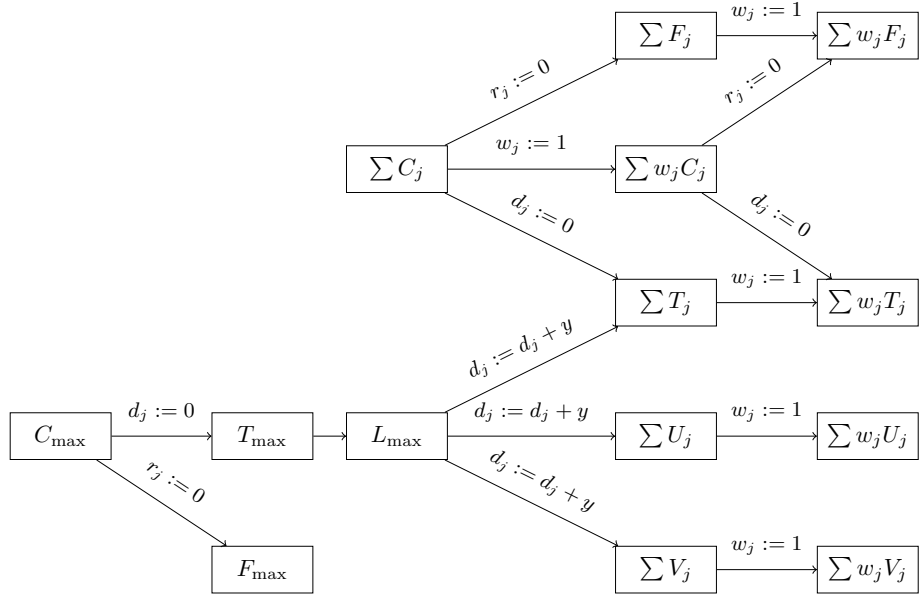

We now revisit classical reductions between the usual objective functions in the context of fine-grained complexity. Reductions like these can e.g. be found in the work by Lawler \etal~\cite{LLRS89}. The content of the following lemma is also visualized in \cref{fig:objectivefunctions}. Moreover, all SETH-based lower bounds -- including those from this section -- are summarized in \cref{tab:results}.

\begin{lemma}\label{lem:objectivefunctions}
Consider machine model $\alpha$ and additional constraints $\beta$. We have:
\begin{enumerate}
    \item If $\gamma=C_{\max}$ and $\gamma'=T_{\max}$ or $\gamma=\sum C_j$ and $\gamma'=\sum T_j$ or $\gamma=\sum w_jC_j$ and $\gamma'=\sum w_jT_j$, there exists a reduction from $\alpha|\beta|\gamma$ to $\alpha|\beta|\gamma'$, where we only introduce zero-due-dates, i.e. $d_j'=0$ for every $j\in[n]$.
    \item If $\gamma=C_{\max}$ and $\gamma'=F_{\max}$ or $\gamma=\sum C_j$ and $\gamma'=\sum F_j$ or $\gamma=\sum w_jC_j$ and $\gamma'=\sum w_jF_j$, there exists a reduction from $\alpha|\beta|\gamma$ to $\alpha|\beta|\gamma'$, where we only introduce zero-release-dates, i.e. $r_j'=0$ for every $j\in[n]$.
    \item If $\gamma=\sum C_j$ and $\gamma'=\sum w_jC_j$ or $\gamma=\sum T_j$ and $\gamma'=\sum w_jT_j$ or $\gamma=\sum U_j$ and $\gamma'=\sum w_jU_j$ or $\gamma=\sum V_j$ and $\gamma'=\sum w_jV_j$ or $\gamma=\sum F_j$ and $\gamma'=\sum w_jF_j$, there exists a reduction from $\alpha|\beta|\gamma$ to $\alpha|\beta|\gamma'$, where we only introduce unit-weights, i.e. $w_j'=1$ for every $j\in[n]$.
    \item If $\gamma=T_{\max}$ and $\gamma'=L_{\max}$, there exists a reduction from $\alpha|\beta|\gamma$ to $\alpha|\beta|\gamma'$, where we do not change anything about the instance.
    \item If $\gamma=L_{\max}$ and $\gamma'=\sum T_j$, $\gamma'=\sum U_j$ or $\gamma'=\sum V_j$, there exists a reduction from $\alpha|\beta|\gamma$ to $\alpha|\beta|\gamma'$, where we only increase the due dates by $y$, i.e. $d_j'=d_j+y$ for every $j\in[n]$ and set the target value to zero, i.e. $y'=0$.
\end{enumerate}
Note that in all of these reductions, only the mentioned parameters are modified. Everything else about the instance (e.g. the number of jobs $n$) stays the same. The running time of each reduction is polynomial in $n$.
\end{lemma}

\begin{proof}
For part 1, we observe that setting the due date of every job $j\in[n]$ to zero means that its completion time $C_j$ is identical to its tardiness $T_j$:
\[T_j=\max\{C_j-d_j,0\}=\max\{C_j,0\}=C_j\]
Hence, minimizing $T_{\max}$, $\sum T_j$ and $\sum w_jT_j$ is equivalent to minimizing $C_{\max}$, $\sum C_j$ and $\sum w_jC_j$, respectively. 

Part 2 is similar: Setting a release date $r_j$ to zero means that the flow time $F_j$ of job $j$ is equal to its completion time $C_j$.

Part 3 is obvious: Weighting all jobs equally is equivalent to having no weights at all. Having the weights all set to 1 also means that the objective value does not change.

For part 4, observe that any schedule that minimizes the maximum lateness $L_{\max}$ also minimizes the maximum tardiness and that we only have non-negative $y$-values in the $T_{\max}$-problem.\footnote{The possibility of negative $y$-values in the $L_{\max}$-problem prevents the same \enquote*{reduction} from working in the other direction.} 

For part 5, observe that the original instance of $\alpha|\beta|\gamma$ has a schedule with maximum lateness $L_{\max}\leq y$ if and only if in the instance with delayed due dates $d_j'=d_j+y$ no job is late. No job being late is equivalent to $\sum T_j$, $\sum U_j$ and $\sum V_j$ all being zero. 

All of the reductions are polynomial in $n$, since we only need to construct new jobs with not too large parameters.
\end{proof}

\begin{xltabular}{\textwidth}{|c|c|c|l|l|}
\hline
$\bm{\alpha}$ & $\bm{\beta}$ & $\bm{\gamma}$ & \textbf{Lower Bound} & \textbf{Ref.} \\\hline
$P2$ & $-$ & $C_{\max}$ & $\Oh{2^{\delta n}(y+P)^{1-\eps}}$ & thm. \ref{thm||C_{max}} \\\hline
$P2$ & $size$ & $C_{\max}$ & $\Oh{2^{\delta n}(y+P)^{1-\eps}}$ & thm. \ref{thm:P2|size|C_{max}} \\\hline
$P2$ & $any$ & $C_{\max}$ & $\Oh{2^{\delta n}(y+P)^{1-\eps}}$ & thm. \ref{thm:P2|any|C_{max}} \\\hline
$P2$ & \makecell[c]{$-$ \\ $size$ \\ $any$} & \makecell[c]{$L_{\max}$ \\ $T_{\max}$} & $\Oh{2^{\delta n}(y+P+f(D))^{1-\eps}}$ & cor. \ref{cor:P2|beta|gamma} \\\hline
$P2$ & \makecell[c]{$-$ \\ $size$ \\ $any$} & \makecell[c]{$\sum U_j$ \\ $\sum V_j$ \\ $\sum T_j$} & $\Oh{2^{\delta n}(d_{\max}+P+f(y))^{1-\eps}}$ & cor. \ref{cor:P2|beta|gamma} \\\hline
$P2$ & \makecell[c]{$-$ \\ $size$ \\ $any$} & \makecell[c]{$\sum w_jU_j$ \\ $\sum w_jV_j$ \\ $\sum w_jT_j$} & $\Oh{2^{\delta n}(d_{\max}+P+f(w_{\max},y))^{1-\eps}}$ & cor. \ref{cor:P2|beta|gamma} \\\hline
$P2$ & \makecell[c]{$-$ \\ $size$ \\ $any$} & $F_{\max}$ & $\Oh{2^{\delta n}(y+P+f(R))^{1-\eps}}$ & cor. \ref{cor:P2|beta|gamma} \\\hline
$P2$ & $-$ & $\sum w_jC_j$ & $\Oh{2^{\delta n}(\sqrt{y}+P+W)^{1-\eps}}$ & thm. \ref{thm||sumw_jC_j} \\\hline
$P2$ & $-$ & $\sum w_jT_j$ & $\Oh{2^{\delta n}(\sqrt{y}+P+W+f(D))^{1-\eps}}$ & cor. \ref{cor:P2||gamma} \\\hline
$P2$ & $-$ & $\sum w_jF_j$ & $\Oh{2^{\delta n}(\sqrt{y}+P+W+f(R))^{1-\eps}}$ & cor. \ref{cor:P2||gamma} \\\hline
$1$ & $Rej\leq Q$ & $C_{\max}$ & $\Oh{2^{\delta n} (y+P+Q+W)^{1-\eps}}$ & thm. \ref{thm:1|Rej|C_{max}} \\\hline
$1$ & $Rej\leq Q$ & \makecell[c]{$L_{\max}$ \\ $T_{\max}$} & $\Oh{2^{\delta n}(y+P+Q+W+f(D))^{1-\eps}}$ & cor. \ref{cor:1|Rej|gamma} \\\hline
$1$ & $Rej\leq Q$ & \makecell[c]{$\sum U_j$ \\ $\sum V_j$ \\ $\sum T_j$} & $\Oh{2^{\delta n}(d_{\max}+P+Q+W+f(y))^{1-\eps}}$ & cor. \ref{cor:1|Rej|gamma} \\\hline
$1$ & $Rej\leq Q$ & \makecell[c]{$\sum w_jU_j$ \\ $\sum w_jV_j$ \\ $\sum w_jT_j$} & $\Oh{2^{\delta n}(d_{\max}+P+Q+W+f(y,\overbar{w}_{\max}))^{1-\eps}}$ & cor. \ref{cor:1|Rej|gamma} \\\hline
$1$ & $Rej\leq Q$ & $F_{\max}$ & $\Oh{2^{\delta n} (y+P+Q+W+f(R))^{1-\eps}}$ & cor. \ref{cor:1|Rej|gamma} \\\hline
$1$ & $-$ & $\sum T_j$ & $\Oh{2^{\delta n}P^{1-\eps}}$ & thm. \ref{thm:1||sumT_j} \\\hline
$1$ & $-$ & $\sum w_jT_j$ & $\Oh{2^{\delta n}(P+f(w_{\max}))^{1-\eps}}$ & cor. \ref{cor:1||sumw_jT_j} \\\hline
$1$ & $-$ & $\sum w_jU_j$ & $\Oh{2^{\delta n} (d_{\max}+y+P+W)^{1 - \eps}}$ & thm. \ref{thm:1||sumw_jU_j} \\\hline
$1$ & $r_j$ & $\sum w_jC_j$ & $\Oh{2^{\delta n}(P+W+r_{\max}+\sqrt{y})^{1-\eps}}$ & thm. \ref{thm:1|r_j|sumw_jC_j} \\\hline
$1$ & $r_j$ & $\sum w_jT_j$ & $\Oh{2^{\delta n}(P+W+r_{\max}+\sqrt{y}+f(D))^{1-\eps}}$ & cor. \ref{cor:1|r_j|sumw_jT_j} \\\hline
$1$ & - & $\sum w_jF_j$ & $\Oh{2^{\delta n} \left(r_{\max}+P+W+\sqrt{y}+f(D)\right)^{1 - \eps}}$ & cor. \ref{cor:1|r_j|sumw_jT_j} \\\hline
$1$ & $r_j$ & $T_{\max}$ & $\Oh{2^{\delta n} (d_{\max} + r_{\max} + P + y)^{1 - \eps}}$ & thm. \ref{thm:1|r_j|T_{max}} \\\hline
$1$ & $r_j$ & $L_{\max}$ & $\Oh{2^{\delta n}(d_{\max}+r_{\max}+P+y)^{1-\eps}}$ & cor. \ref{cor:1|r_j|gamma} \\\hline
$1$ & $r_j$ & \makecell[c]{$\sum U_j$ \\ $\sum V_j$ \\ $\sum T_j$} & $\Oh{2^{\delta n}(d_{\max}+r_{\max}+P+f(y))^{1-\eps}}$ & cor. \ref{cor:1|r_j|gamma} \\\hline
$1$ & $r_j$ & \makecell[c]{$\sum w_jU_j$ \\ $\sum w_jV_j$ \\ $\sum w_jT_j$} & $\Oh{2^{\delta n}(d_{\max}+r_{\max}+P+f(y,w_{\max}))^{1-\eps}}$ & cor. \ref{cor:1|r_j|gamma} \\\hline
$1$ & $-$ & $\sum w_jT_j$ & $\Oh{2^{\delta n}(P+d_{\max}+W+\sqrt{y})^{1-\eps}}$ & thm. \ref{thm:1||sumw_jT_j} \\\hline
$1$ & $C_j\leq d_j$ & $\sum w_jC_j$ & $\Oh{2^{\delta n}(P+d_{\max}+W+\sqrt{y})^{1-\eps}}$ & thm. \ref{thm:1|d_j|sumw_jC_j} \\\hline
$1$ & $C_j\leq d_j$ & $\sum w_jF_j$ & $\Oh{2^{\delta n}(P+d_{\max}+W+\sqrt{y}+f(R))^{1-\eps}}$ & cor. \ref{cor:1|d_j|sumw_jF_j} \\\hline
$1$ & $C_j\leq d_j$ & $\sum w_jT_j$ & $\Oh{2^{\delta n}(P+d_{\max}+W+\sqrt{y}+f(D))^{1-\eps}}$ & cor. \ref{cor:1|d_j|sumw_jF_j} \\\hline
\caption{Overview of our SETH-based lower bounds. Throughout, $f$ is some arbitrary computable function.}
\label{tab:results}
\end{xltabular}

In \cref{thm:1|Rej|C_{max}}, we showed a lower bound for \tf{\one}{\rej}{\cmax}. Using~\cref{lem:objectivefunctions}, we now transfer this result to other rejection problems with more difficult objective functions:
\begin{corollary}\label{cor:1|Rej|gamma}
  Let $f$ be some computable function, $\gamma_1\in\{L_{\max},T_{\max}\}$, $\gamma_2\in\{\sum U_j, \sum V_j,$ $\sum T_j\}$ and $\gamma_3\in\{\sum w_jU_j, \sum w_jV_j, \sum \overbar{w}_jT_j\}$.\footnote{Note that in the case of weighted objective functions we already have weights for each job (i.e. rejection penalties). In order for the reduction to work, we need the weights for the objective function to be conceptually different weights $\overbar{w}_j$.} Then unless SETH fails, for every $\eps>0$ there exists a $\delta>0$ such that there is no algorithm solving
  \begin{enumerate}
      \item \tf{\one}{\rej}{\gamma_1} in time $\Oh{2^{\delta n}(y+P+Q+W+f(D))^{1-\eps}}$,
      \item \tf{\one}{\rej}{\gamma_2} in time $\Oh{2^{\delta n}(d_{\max}+P+Q+W+f(y))^{1-\eps}}$,
      \item \tf{\one}{\rej}{\gamma_3} in time $\Oh{2^{\delta n}(d_{\max}+P+Q+W+f(y,\overbar{w}_{\max}))^{1-\eps}}$ or
      \item \tf{\one}{\rej}{\fmax} in time $\Oh{2^{\delta n} (y+P+Q+W+f(R))^{1-\eps}}$.
  \end{enumerate}
\end{corollary}

\begin{proof}
The lower bound for \tf{\one}{\rej}{\cmax} from \cref{thm:1|Rej|C_{max}} for \tf{\one}{\rej}{\cmax} is $\Oh{2^{\delta n} (y+P+Q+W)^{1-\eps}}$. We use the reductions from \cref{lem:objectivefunctions} to obtain lower bounds for the more difficult objective functions.

For part 1, the reductions only introduce zero-due-dates. Hence, $D=0$ and we get the lower bound $\Oh{2^{\delta n}(y+P+Q+W+f(D))^{1-\eps}}$, where $f$ is some computable function that only depends on $D$ and is hence constant for $D=0$.

For part 2, applying the reductions in sequence gives us due dates that are equal to the original target, together with a new target $y=0$. Hence, we get the lower bound $\Oh{2^{\delta n}(d_{\max}+P+Q+W+f(y))^{1-\eps}}$.

For the third part, we get $\Oh{2^{\delta n}(d_{\max}+P+Q+W+f(y,\overbar{w}_{\max}))^{1-\eps}}$ as our lower bound, since the reduction only adds unit-weights.

In the case of $F_{\max}$, we simply introduce zero-release-dates and get the lower bound $\Oh{2^{\delta n} (y+P+Q+W+f(R))^{1-\eps}}$.
\end{proof}

Similarly, the lower bounds from \cref{thm||C_{max}}, \cref{thm:P2|size|C_{max}} and \cref{thm:P2|any|C_{max}} for the two-machine problems with $C_{\max}$-objective can also be transferred to problems with more difficult objective functions:
\begin{corollary}\label{cor:P2|beta|gamma}
  Let $f$ be some computable function, $\beta\in\{size, any, -\}$, $\gamma_1\in\{L_{\max},T_{\max}\}$, $\gamma_2\in\{\sum U_j, \sum V_j, \sum T_j\}$ and $\gamma_3\in\{\sum w_jU_j, \sum w_jV_j, \sum w_jT_j\}$. Then unless SETH fails, for every $\eps>0$ there exists a $\delta>0$ such that there is no algorithm solving
  \begin{enumerate}
      \item \tf{\two}{\beta}{\gamma_1} in time $\Oh{2^{\delta n}(y+P+f(D))^{1-\eps}}$,
      \item \tf{\two}{\beta}{\gamma_2} in time $\Oh{2^{\delta n}(d_{\max}+P+f(y))^{1-\eps}}$,
      \item \tf{\two}{\beta}{\gamma_3} in time $\Oh{2^{\delta n}(d_{\max}+P+f(w_{\max},y))^{1-\eps}}$ or
      \item \tf{\two}{\beta}{\fmax} in time $\Oh{2^{\delta n}(y+P+f(R))^{1-\eps}}$.
  \end{enumerate}
\end{corollary}
\begin{proof}
The lower bounds for \tf{\two}{\beta}{\cmax} are all $\Oh{2^{\delta n}(y+P)^{1-\eps}}$. We apply \cref{lem:objectivefunctions} and add the additional parameters to the lower bound.
  
In case 1 and 4, the reduction from \tf{\two}{\beta}{\cmax} to \tf{\two}{\beta}{\gamma_1} (resp. \tf{\two}{\beta}{\fmax}) involves only adding zero-due-dates (resp. zero-release-dates). Hence, $D=R=0$ in the constructed instances and the lower bounds follow.
  
In case 2, the instance is constructed by adding due dates equal to the original target value and the new target value is set to zero. So the new lower bound becomes $\Oh{2^{\delta n}(d_{\max}+P+f(y))^{1-\eps}}$.
  
In case 3, in addition to due dates, we get unit-weights. So we get the lower bound $\Oh{2^{\delta n}(d_{\max}+P+f(w_{\max},y))^{1-\eps}}$, concluding the proof.
\end{proof}

The lower bound for \tf{\two}{}{\sumwc} from~\cref{thm||sumw_jC_j}, together with~\cref{lem:objectivefunctions} yields the following implications:
\begin{corollary}\label{cor:P2||gamma}
  Let $f$ be any computable function. Then unless SETH fails, for every $\eps>0$ there exists a $\delta>0$ such that there is no algorithm solving
  \begin{enumerate}
      \item \tf{\two}{}{\sumwt} in time $\Oh{2^{\delta n}(\sqrt{y}+P+W+f(D))^{1-\varepsilon}}$ or
      \item \tf{\two}{}{\sumwf} in time $\Oh{2^{\delta n}(\sqrt{y}+P+W+f(R))^{1-\varepsilon}}$.
  \end{enumerate}
\end{corollary}

\begin{proof}
We have $\Oh{2^{\delta n} (\sqrt{y}+P+W)^{1-\eps}}$ as our lower bound for \tf{\two}{}{\sumwc} (see \cref{thm||sumw_jC_j}) and we use the reductions from \cref{lem:objectivefunctions}.

In both cases, we only introduce zero-due-dates/zero-release-dates and have $D=R=0$; so the lower bounds follow.
\end{proof}

Using the lower bound for \tf{\one}{}{\sumt} from \cref{thm:1||sumT_j} and \cref{lem:objectivefunctions}, we get the following result by adding unit-weights:
\begin{corollary}\label{cor:1||sumw_jT_j}
  Let $f$ be any computable function. Then for every $\eps>0$, there is a $\delta>0$ such that \tf{\one}{}{\sumwt} cannot be solved in time $\Oh{2^{\delta n}(P+f(w_{\max}))^{1-\eps}}$, unless the SETH fails.
\end{corollary}

\begin{proof}
The lower bound for \tf{\one}{}{\sumt} from \cref{thm:1||sumT_j} is $\Oh{2^{\delta n} P^{1-\eps}}$ and using the unit-weight reduction from \cref{lem:objectivefunctions}, we get the claimed lower bound, since $w_{\max}=1$.
\end{proof}

As noted before, \tf{\one}{\rj}{\sumwc} is equivalent to \tf{\one}{}{\sumwf}. Hence, we get the same lower bound for that problem. By introducing zero-due-dates, we also get a lower bound for \tf{\one}{\rj}{\sumwt} using \cref{lem:objectivefunctions}:
\begin{corollary}\label{cor:1|r_j|sumw_jT_j}
	Let $f$ be any computable function. Then unless the SETH fails, for every $\eps>0$, there exists a $\delta>0$ such that there is no algorithm solving
	\begin{enumerate}
	    \item \tf{\one}{\rj}{\sumwt} in time $\Oh{2^{\delta n} \left(r_{\max}+P+W+\sqrt{y}+f(D)\right)^{1 - \eps}}$ or
	    \item \tf{\one}{}{\sumwf} in time $\Oh{2^{\delta n} \left(r_{\max}+P+W+\sqrt{y}\right)^{1 - \eps}}$.
	\end{enumerate}
\end{corollary}

\begin{proof}
From \cref{thm:1|r_j|sumw_jC_j}, we have $\Oh{2^{\delta n} \left(r_{\max}+P+W+\sqrt{y}\right)^{1 - \eps}}$ as lower bound for \tf{\one}{\rj}{\sumwc}, so using the zero-due-dates reduction from \cref{lem:objectivefunctions}, we get $D=0$ and the lower bound for \tf{\one}{\rj}{\sumwt} follows.
\end{proof}

Using our lower bound for \tf{\one}{\rj}{\tmax} from \cref{thm:1|r_j|T_{max}} and \cref{lem:objectivefunctions}, we get the following implications for other single-machine release date problems:
\begin{corollary}\label{cor:1|r_j|gamma}
  Let $\gamma_1\in\{\sum U_j, \sum V_j, \sum T_j\}$, $\gamma_2\in\{\sum w_jU_j, \sum w_jV_j, \sum w_jT_j\}$ and let $f$ be any computable function. Then unless SETH fails, for every $\eps>0$ there exists a $\delta>0$ such that there is no algorithm solving
  \begin{enumerate}
      \item \tf{\one}{\rj}{\lmax} in time $\Oh{2^{\delta n} (d_{\max} + r_{\max} + P + y)^{1 - \eps}}$,
      \item \tf{\one}{\rj}{\gamma_1} in time $\Oh{2^{\delta n} (d_{\max} + r_{\max} + P + f(y))^{1 - \eps}}$ or
      \item \tf{\one}{\rj}{\gamma_2} in time $\Oh{2^{\delta n} (d_{\max} + r_{\max} + P + f(y,w_{\max}))^{1 - \eps}}$.
  \end{enumerate}
\end{corollary}
\begin{proof}
We have the lower bound $\Oh{2^{\delta n} (d_{\max} + r_{\max} + P + y)^{1 - \eps}}$ for \tf{\one}{\rj}{\tmax} (see \cref{thm:1|r_j|T_{max}}) and using the reductions from \cref{lem:objectivefunctions}, we get the following results:

The lower bound for \tf{\one}{\rj}{\lmax} directly follows, since the reduction does not change the instance in any way.

For the second part, the reduction increases the due dates by the original target value and sets the new target value to zero. Hence, the $\Oh{2^{\delta n} (d_{\max} + r_{\max} + P + f(y))^{1 - \eps}}$-time lower bound follows.

The same lower bound holds for the third case, but we also introduce unit-weights, so the bound becomes $\Oh{2^{\delta n} (d_{\max} + r_{\max} + P + f(y,w_{\max}))^{1 - \eps}}$.
\end{proof}

Using \cref{lem:objectivefunctions}, we can also transfer the lower bound for \tf{\one}{\dj}{\sumwc} from \cref{thm:1|d_j|sumw_jC_j} to weighted flow minimization and total weighted tardiness.
\begin{corollary}\label{cor:1|d_j|sumw_jF_j}
	Let $f$ be any computable function. Then unless the SETH fails, for every $\eps>0$, there exists a $\delta>0$ such that there is no algorithm solving
	\begin{enumerate}
	    \item \tf{\one}{\dj}{\sumwf} in time $\Oh{2^{\delta n} \left(d_{\max}+P+W+\sqrt{y}+f(R)\right)^{1 - \eps}}$ or
	    \item \tf{\one}{\dj}{\sumwt} in time $\Oh{2^{\delta n} \left(d_{\max}+P+W+\sqrt{y}+f(D)\right)^{1 - \eps}}$.
	\end{enumerate}
	Here, $D$ is the sum of due dates and $d_{\max}$ is the largest deadline.
\end{corollary}

\begin{proof}
From \cref{thm:1|d_j|sumw_jC_j}, we have $\Oh{2^{\delta n} \left(d_{\max}+P+W+\sqrt{y}\right)^{1 - \eps}}$ as a lower bound for \tf{\one}{\dj}{\sumwc}. The lower bounds for \tf{\one}{\dj}{\sumwf} and \tf{\one}{\dj}{\sumwt} follow if we use the zero-release-dates/zero-due-dates reduction from \cref{lem:objectivefunctions}. 
\end{proof}

Finally, we can transfer the result for \tf{\fixed}{}{\cmax} from \cref{thm:fixed} to other objectives, using the reductions from \cref{lem:objectivefunctions}:
\begin{corollary}\label{cor:fixed}
  Let $\gamma\in\{F_{\max},T_{\max},L_{\max},\sum T_j,\sum U_j,\sum V_j,\sum w_jT_j,\sum w_jU_j,$ $\sum w_jV_j\}$. There is no algorithm solving \tf{\fixed}{}{\gamma} in time $\Oh{nmP^{\oh{\frac{m}{\log^2(m)}}}}$, unless the ETH fails.
\end{corollary}
\begin{proof}
Here, we use simply that there is a reduction from \tf{\fixed}{}{\cmax} to each of these problems, where neither the number of jobs, nor the number of machines or the processing times change. The lower bound then directly holds for the more difficult problems.
\end{proof}
\end{document}